\documentclass[10pt,twocolumn,english,showpacs,preprintnumbers,amsmath,amssymb,floatfix,nofootinbib]{revtex4-1} 

\usepackage[normalem]{ulem}    
\usepackage{caption}
\usepackage{tikz,xcolor}
\usepackage[colorlinks = true,
linkcolor = blue,
urlcolor  = blue,
citecolor = blue,
anchorcolor = blue]{hyperref}

\definecolor{lime}{HTML}{A6CE39}
\DeclareRobustCommand{\orcidicon}{%
	\begin{tikzpicture}
		\draw[lime, fill=lime] (0,0)
		circle [radius=0.16]
		node[white] {{\fontfamily{qag}\selectfont \tiny ID}};
		\draw[white, fill=white] (-0.0625,0.095)
		circle [radius=0.007];
	\end{tikzpicture}
	\hspace{-2mm}
}
\foreach \x in {A, ..., Z}{%
\expandafter\xdef\csname orcid\x\endcsname{\noexpand\href{https://orcid.org/\csname orcidauthor\x\endcsname}{\noexpand\orcidicon}}
}

\usepackage{amsmath}
\usepackage[T1]{fontenc}
\usepackage[latin9]{inputenc}
\usepackage{color}
\usepackage{array}
\usepackage{amstext}
\usepackage{subfig}
\usepackage{graphicx}
\usepackage{esint}
\usepackage{rotating}
\usepackage{slashed}
\usepackage{siunitx}
\usepackage[normalem]{ulem}
\usepackage[font=small,labelfont=bf]{caption}
\usepackage{lipsum}
\usepackage{xcolor}
\usepackage{comment}
\usepackage{amssymb}

\usepackage{booktabs}

\newcommand{\xpom}{x_{\xpom}}

\makeatletter

\@ifundefined{textcolor}{}
{%
 \definecolor{BLACK}{gray}{0}
 \definecolor{WHITE}{gray}{1}
 \definecolor{RED}{rgb}{1,0,0}
 \definecolor{GREEN}{rgb}{0,1,0}
 \definecolor{BLUE}{rgb}{0,0,1}
 \definecolor{CYAN}{cmyk}{1,0,0,0}
\definecolor{MAGENTA}{cmyk}{0,1,0,0}
 \definecolor{YELLOW}{cmyk}{0,0,1,0}
 }

\@ifundefined{definecolor}
{\usepackage{color}}{}
\@ifundefined{definecolor}
{\usepackage{color}}{}
\makeatother
\usepackage{babel}

\def\Re{{\cal R \mskip-4mu \lower.1ex \hbox{\it e}\,}}
\def\Im{{\cal I \mskip-5mu \lower.1ex \hbox{\it m}\,}}

\def\tev{\,{\ifmmode\mathrm {TeV}\else TeV\fi}}
\def\gev{\,{\ifmmode\mathrm {GeV}\else GeV\fi}}
\def\mev{\,{\ifmmode\mathrm {MeV}\else MeV\fi}}

\begin{document} 

%
%
\title{Toward Precision Helicity PDFs from Global DIS and SIDIS Fits with Projected EIC Measurements}

\collaboration{HAPS Collaboration}

%
%

\author{Hamzeh~Khanpour$^{1,2}$\orcidE{}}
\email{Hamzeh.Khanpour@cern.ch}

\author{Maryam~Soleymaninia$^{2}$\orcidB{}}
\email{Maryam\_Soleymaninia@ipm.ir}

\author{Majid~Azizi$^{2}$\orcidD{}}
\email{Ma.Azizi@ipm.ir}

\author{Michael~Klasen$^{3}$\orcidG{}}
\email{Michael.Klasen@uni-muenster.de}

\author{Hadi~Hashamipour$^{4}$\orcidA{}}
\email{Hadi.Hashamipour@lnf.infn.it}

\author{Maral~Salajegheh$^{5}$\orcidC{}} 
\email{Maral@hiskp.uni-bonn.de }

\author{Ulf-G.~Mei{\ss}ner$^{5,6,7}$\orcidF{}}
\email{Meissner@hiskp.uni-bonn.de}

\affiliation {
$^{1}$AGH University, Faculty of Physics and Applied Computer Science, Al. Mickiewicza 30, 30-055 Krakow, Poland.                             \\  
$^{2}$School of Particles and Accelerators, Institute for Research in Fundamental Sciences (IPM), P.O.Box 19395-5531, Tehran, Iran.           \\
$^{3}$Institut f\"ur Theoretische Physik, Universit\"at M\"unster, \\ Wilhelm-Klemm-Stra\ss{}e 9, 48149 M\"unster, Germany.                   \\ 
$^{4}$Istituto Nazionale di Fisica Nucleare, Gruppo collegato di Cosenza, I-87036 Arcavacata di Rende, Cosenza, Italy.                        \\  
$^{5}$Helmholtz-Institut f\"ur Strahlen-und Kernphysik and Bethe Center for Theoretical Physics, Universit\"at Bonn, D-53115 Bonn, Germany.   \\  
$^{6}$Institute for Advanced Simulation (IAS-4), Forschungszentrum J\"ulich, D-52425 J\"ulich, Germany.                                       \\
$^{7}$Peng Huanwu Collaborative Center for Research and Education, International Institute for Interdisciplinary and Frontiers, Beihang University, Beijing 100191, China                                                                                       \\
}

\date{\today}

%
\begin{abstract}

We present HAPS-pPDF1.0, a new global determination of the helicity-dependent parton distribution functions (PDFs) of the proton,
based on inclusive deep-inelastic scattering (DIS) and semi-inclusive DIS (SIDIS) data within a consistent
next-to-leading order (NLO) QCD framework. 
In addition to existing measurements, we incorporate simulated pseudodata for the future Electron-Ion Collider (EIC),
considering two beam-energy configurations, $E_e \times E_p = 5 \times 41~\mathrm{GeV^2}$ and $18 \times 275~\mathrm{GeV^2}$,
corresponding to an extended kinematic reach down to $x \sim 10^{-5}$. 
We focus on longitudinal double-spin asymmetries $A_1^h$ for charge-separated pion and kaon production in SIDIS
off a longitudinally polarized proton target.  
These projected measurements significantly improve the flavor separation of sea-quark polarized PDFs
($\Delta \bar{u}$, $\Delta \bar{d}$, $\Delta s$) and reduce the uncertainties on 
both quark and gluon helicity distributions, with the largest impact at small $x$. 
Polarized PDFs are extracted using a neural-network  parametrization and a Monte Carlo  replica methodology to propagate
experimental uncertainties, while theoretical constraints such as positivity are imposed during the fit. 
We demonstrate that the inclusion of EIC pseudodata leads to a substantially more precise determination of polarized PDFs,
with the largest impact in the small-$x$ region. This analysis is performed within the publicly available \textsc{MontBlanc} framework, and the resulting polarized PDF sets are provided in the standard \textsc{LHAPDF} format. 

\end{abstract}

\maketitle
\tableofcontents{}

\section{Introduction}\label{Introduction}

A precise understanding of the internal spin structure of the proton remains a central goal of
high-energy nuclear and particle physics. Within QCD, the proton spin can be decomposed into contributions
from quark and gluon helicities and from partonic orbital angular momentum. 
The experimental discovery that quark spins account for only a limited fraction of the proton spin,
first observed in polarized inclusive deep-inelastic scattering (DIS) measurements of the spin-dependent
structure function $g_1$ by the European Muon Collaboration (EMC), initiated the so-called
\textit{proton spin puzzle} and motivated an extensive experimental and theoretical program over the past
three decades~\cite{EuropeanMuon:1987isl,EuropeanMuon:1989yki,Ji:2020ena}. 

A key ingredient of this program is the determination of helicity-dependent (polarized, henceforth) 
parton distribution functions (PDFs), which quantify the difference between the densities of partons with
spin aligned parallel and anti-parallel to the longitudinal polarization of the parent proton. 
They are denoted as $\Delta q(x,Q^2)$ for quarks and $\Delta g(x,Q^2)$ for gluons, and their lowest moments
are related to axial currents and to the spin decomposition of the proton~\cite{Anselmino:1994gn,Ji:2020ena,Ethier:2020way}. 
In contrast to unpolarized PDFs, the polarized PDFs can only be constrained through measurements of
spin-dependent observables, most notably inclusive DIS asymmetries and polarized structure functions,
and semi-inclusive DIS (SIDIS) asymmetries in which identified hadrons are detected in the final state. 
In SIDIS, the sensitivity to individual flavors is enhanced by the use of fragmentation functions (FFs),
which connect partonic flavors to charge-separated hadron yields. 

Over the years, global QCD analyses of helicity PDFs have incorporated an increasingly diverse set of data,
including inclusive DIS measurements from fixed-target experiments and, more recently,
longitudinal spin asymmetries from polarized proton-proton collisions at RHIC, which provide essential 
constraints on the gluon helicity distribution $\Delta g(x,Q^2)$. 
Modern global determinations differ in their fitted data sets, perturbative accuracy, 
and methodology~~\cite{Gluck:2000dy,Hirai:2003pm,Hirai:2006sr,Blumlein:2010rn,Leader:2010rb,Adamiak:2023yhz,deFlorian:2009vb,Nocera:2014uea,deFlorian:2008mr,deFlorian:2014yva,
Sato:2016tuz,Ethier:2017zbq,Nocera:2014gqa,Bertone:2024taw,Borsa:2024mss,Cruz-Martinez:2025ahf,Arbabifar:2024ftq,Khanpour:2017fey,Khanpour:2017cha,Borsa:2020lsz,Cocuzza:2025qvf,JAMCollaborationSmall-xAnalysisGroup:2025tfa,Borsa:2024mss,Bonino:2025bqa,Goyal:2024emo,Bonino:2024rku,Gehrmann:2025xab}.
While inclusive DIS alone has limited sensitivity to the separation of quark and antiquark helicities and
to $\Delta g$, the combined use of DIS, SIDIS, and (where included) proton-proton data enables a more
differential flavor decomposition and a stronger handle on the gluon contribution. 
Recent years have also seen major progress in the theoretical description of polarized observables,
including higher-order QCD corrections to inclusive DIS structure functions 
and to SIDIS structure functions~\cite{Borsa:2024mss,Bonino:2025bqa,Goyal:2024emo,Bonino:2024rku,Gehrmann:2025xab},
as well as improvements in PDF-fitting methodologies based on neural-network parameterization and Monte Carlo
representations of uncertainties~\cite{Bertone:2024taw,Cruz-Martinez:2025ahf}. 

On the experimental side, the forthcoming Electron-Ion Collider (EIC) is expected to provide a qualitative leap in our knowledge of
polarized PDFs~\cite{AbdulKhalek:2021gbh,AbdulKhalek:2022hcn,Borsa:2020lsz}, proton PDFs~\cite{Azizi:2024swj,Armesto:2023hnw,Jimenez-Lopez:2026bxi} and fragmentation functions (FFs)~\cite{Soleymaninia:2025cvi,Aschenauer:2019kzf}. 
The EIC will collide longitudinally polarized lepton and proton beams with high luminosity and broad
kinematic coverage in the proton momentum fraction $x$ and the hard scale $Q^2$. 
This will enable percent-level measurements of inclusive and semi-inclusive spin asymmetries over a range
that significantly extends that of existing data (see, \textit{e.g.}, the EIC projections discussed
in Refs.~\cite{Ethier:2020way,AbdulKhalek:2021gbh,Borsa:2020lsz}). 
In particular, access to the small-$x$ region is essential because present constraints on $\Delta g$
and on sea-quark polarized PDFs remain comparatively weak there, and the corresponding contributions to the
proton spin are therefore still uncertain.

A distinctive asset of the EIC program is the measurement of longitudinal double-spin asymmetries for
charge-separated pion and kaon production in SIDIS process. 
These observables provide enhanced sensitivity to the flavor structure of the polarized sea,
including $\Delta\bar u$, $\Delta\bar d$, and $\Delta s$, through the combined dependence on helicity PDFs
and FFs~\cite{Borsa:2020lsz}.  
Quantifying the impact of these measurements requires realistic projections that account for detector
acceptance, binning, and systematic effects. 
In this work we adopt the EIC pseudodata generation strategy developed in the ECCE studies of
Ref.~\cite{Adkins:2022jfp} and, specifically, we follow the detector-level pseudodata framework described in
Ref.~\cite{VanHulse:2023uga}, which provides projected SIDIS asymmetries for charge-separated pions and kaons
for two beam-energy configurations, $E_e \times E_p = 5 \times 41~\mathrm{GeV^2}$  and $18 \times 275~\mathrm{GeV^2}$,
corresponding to an extended kinematic reach down to $x \sim 10^{-5}$. 

The main aim of the present paper is to provide a new global determination of polarized PDFs based on a consistent
next-to-leading order (NLO) QCD framework for polarized inclusive DIS, SIDIS, and DGLAP evolution, and to quantify the impact of
EIC projected measurements. 
Our analysis closely follows the methodology developed in the MAP Collaboration determination of polarized PDFs,
notably the use of a neural-network parametrization and a Monte Carlo replica representation of uncertainties,
together with theoretical constraints such as positivity~\cite{Bertone:2024taw}. 
Building upon this baseline, we include EIC pseudodata for SIDIS longitudinal asymmetries in pion and kaon
production, generated following Ref.~\cite{VanHulse:2023uga}, and we assess their impact on the precision
of sea-quark flavor separation and on the small-$x$ behavior of polarized PDFs. 

Our baseline fit is intentionally restricted to polarized inclusive DIS and SIDIS data, 
so as to isolate in a controlled way the incremental impact of projected EIC SIDIS measurements within the same DIS/SIDIS framework. 
Hence, the polarized proton-proton measurements from RHIC, such as jet and $W$-boson asymmetry data, are not 
included here, although they provide important complementary constraints, 
especially on $\Delta g$ and the sea-quark flavor decomposition, and their inclusion 
is left for a future extended global analysis. 

The primary objectives of this study are: 
(i) to improve the determination of quark and gluon polarized PDFs by incorporating EIC pseudodata,
with particular emphasis on the small-$x$ region; and 
(ii) to strengthen the flavor separation of the polarized sea using charge-separated SIDIS measurements,
thereby enabling tighter constraints on $\Delta\bar u$, $\Delta\bar d$, and the strange sector.

The structure of this paper is as follows.
In Sec.~\ref{Theoretical_predictions} we summarize the theoretical framework used to compute DIS and SIDIS
observables at NLO. 
The experimental input and the EIC pseudodata sets are described in Sec.~\ref{data_sets}. 
The fitting methodology, including the neural-network parametrization and Monte Carlo replica strategy,
is presented in this section as well.  
Results are discussed in Sec.~\ref{Results}, where we compare fits with and without EIC pseudodata and 
quantify the resulting improvements in flavor separation and small-$x$ constraints. 
Finally, conclusions and an outlook for future research directions in this area are given in Sec.~\ref{summary}.

\section{Theoretical input}\label{Theoretical_predictions}

In this work, the theoretical framework for polarized inclusive DIS and SIDIS observables is computed
consistently at NLO in perturbative QCD within leading-twist collinear factorization. 
Polarized PDFs are parametrized at the input scale $Q_0^2=1~\mathrm{GeV}^2$ and evolved to higher scales
using the NLO DGLAP evolution equations for spin-dependent parton distributions. 
All results are obtained in the $\overline{\mathrm{MS}}$ renormalization and factorization schemes. 
We work in a zero-mass variable-flavor-number scheme (ZM-VFNS) and therefore neglect heavy-quark mass effects~\cite{Hekhorn:2018ywm,Behring:2015zaa,Ablinger:2019etw,Behring:2021asx,Blumlein:2021xlc,Bierenbaum:2022biv,Ablinger:2023ahe}; in practice, this choice is adequate for the kinematics of the data sets
considered here (see also Ref.~\cite{Bertone:2024taw,Hekhorn:2024tqm} for a recent discussion).

\paragraph{Inclusive DIS:}

The experimental information entering the fit is provided in terms of the spin-dependent structure function
$g_1(x,Q^2)$ (or ratios involving $g_1$), which in leading-twist collinear factorization can be written as~\cite{Blumlein:2010rn,Leader:2010rb,Adamiak:2023yhz,deFlorian:2009vb,Nocera:2014uea}
%
\begin{align}
g_1(x,Q^2)
&= \frac{1}{2}\, \sum_{q} e_q^2
\Big[
\Delta q(x,Q^2)\,\otimes_x\,\Delta C_q(x,Q^2)
\nonumber\\
&\hspace{18mm}
+
\Delta g(x,Q^2)\,\otimes_x\,\Delta C_g(x,Q^2)
\Big]\,,
\label{eq:g1_DIS}
\end{align}
%

where the sum runs over active quark and antiquark flavors, $e_q$ is the electric charge of flavor $q$; 
$\Delta q$ and $\Delta g$ denote quark and gluon polarized PDFs, and $\Delta C_{q,g}$ are the corresponding
massless coefficient functions computed through the NLO in perturbative QCD~\cite{Zijlstra:1993sh}. 
The helicity-dependent splitting functions (polarized anomalous dimensions) entering DGLAP evolution at NLO
follow Refs.~\cite{Mertig:1995ny,Vogelsang:1995vh}.
Polarized splitting functions are also known beyond NLO, including the three-loop (NNLO) results~\cite{Moch:2014sna,Blumlein:2021ryt,Blumlein:2021enk};
however, in the present analysis we consistently use NLO evolution.
The Mellin convolution acting on the Bjorken variable $x$ is defined as
%
\begin{equation}
\left[f\otimes_x g\right](x) \equiv \int_x^1 \frac{dy}{y}\, f(y)\, g\!\left(\frac{x}{y}\right).
\label{eq:convx}
\end{equation}

\paragraph{Semi-inclusive DIS:}

For SIDIS, we consider the longitudinally polarized structure function $g_1^h(x,z,Q^2)$ associated with
the production of an identified hadron $h$ in the current fragmentation region. 
At leading twist, it factorizes into polarized PDFs, perturbative 
coefficient functions, and unpolarized FFs as~\cite{Adamiak:2023yhz}
%
\begin{align}
g_1^{h}(x,z,Q^2)
&= \frac{1}{2}\, \sum_{q} e_q^2
\Bigg\{
\Big[
\Delta q(x,Q^2)\,\otimes_x\,\Delta C_{qq}(x,z,Q^2)
\nonumber\\
&\hspace{-6mm}
+
\Delta g(x,Q^2)\,\otimes_x\,\Delta C_{qg}(x,z,Q^2)
\Big]\otimes_z D_q^{h}(z,Q^2)
\nonumber\\
&\hspace{-6mm}
+
\Delta q(x,Q^2)\,\otimes_x\,\Delta C_{gq}(x,z,Q^2)\,\otimes_z D_g^{h}(z,Q^2)
\Bigg\}\,,
\label{eq:g1_SIDIS}
\end{align}
%
where $z$ denotes the light-cone momentum fraction of the observed hadron with respect to the fragmenting
parton, and $D_i^h(z,Q^2)$ are the corresponding FFs. 
The NLO massless SIDIS coefficient functions entering Eq.~\eqref{eq:g1_SIDIS} are taken from
Refs.~\cite{Furmanski:1981cw,deFlorian:1997zj}. 
The convolution acting on $z$ is defined by
%
\begin{equation}
\left[f\otimes_z g\right](z) \equiv \int_z^1 \frac{d\zeta}{\zeta}\, f(\zeta)\, g\!\left(\frac{z}{\zeta}\right).
\label{eq:convz}
\end{equation}
%
In the present QCD analysis, we use the \texttt{MAPFF1.0} FFs for charged pions and kaons at the NLO
accuracy~\cite{Khalek:2021gxf,AbdulKhalek:2022laj}.

\paragraph{Spin asymmetries:}

The longitudinal double-spin asymmetry for SIDIS is defined as
%
\begin{equation}
A_1^{h}(x,z,Q^2) = \frac{g_1^{h}(x,z,Q^2)}{F_1^{h}(x,z,Q^2)}\,,
\label{eq:A1h_def}
\end{equation}
%
where $F_1^h$ is the corresponding unpolarized SIDIS structure function, built from the unpolarized PDFs and FFs~\cite{Adamiak:2023yhz}, 
%
\begin{align}
F_1^h(x,z,Q^2) &= \frac{1}2 \sum_q e_q^2 \bigg[
q(x,Q^2) \otimes C_q(x,z,Q^2) \otimes D_q^h(z,Q^2) \notag \\
&\quad + g(x,Q^2) \otimes C_g(x,z,Q^2) \notag \\
&\quad\quad \otimes D_g^h(z,Q^2)
\bigg]
\label{eq:F1h_def}
\end{align}
%
It is computed consistently by replacing $\Delta f_i$ and $\Delta C$ in Eq.~\eqref{eq:g1_SIDIS} with their
unpolarized counterparts $f_i$ and $C$. 
The unpolarized  PDFs entering Eq.~\eqref{eq:F1h_def} are taken from the very recent 
\texttt{NNPDF4.0} PDF set~\cite{NNPDF:2021njg} parton set with perturbative charm.
Analogous expressions are used for inclusive DIS observables (such as $g_1/F_1$) whenever 
experiments provide data in ratio form.

\paragraph{Kinematic cuts and nonperturbative corrections:}
To limit the impact of kinematic regions where nonperturbative corrections may become more important, we impose the cuts
$Q^2 > 1~\mathrm{GeV}^2$ and $W^2 = Q^2(1-x)/x > 4~\mathrm{GeV}^2$ on all fitted data points.
These cuts reduce, but do not eliminate, sensitivity to target-mass corrections (TMCs)~\cite{Blumlein:1998nv,Schienbein:2007gr,Khanpour:2017cha}, higher-twist contributions (HT)~\cite{Leader:2009tr,Khanpour:2017cha,Cocuzza:2025qvf}, 
and nuclear effects in deuteron data. 
Accordingly, we do not include TMC, HT, or deuteron nuclear corrections in the present analysis, and our results should be interpreted as a leading-twist NLO determination within this approximation. 
The practical justification for this choice is not that such effects are absent, but rather that, within the scope and precision of the present study, a leading-twist treatment provides an adequate description of the fitted data.
A dedicated assessment of these nonperturbative corrections is left for future work.

\paragraph{Implementation:}

All the expressions for the coefficient and splitting functions are 
implemented in the publicly available code \texttt{APFEL++}~\cite{Bertone:2013vaa,Bertone:2017gds}, 
which one can use to compute the theoretical predictions that enter the fit. 
All theoretical calculations - PDF evolution, polarized inclusive DIS and SIDIS structure functions, and the corresponding
spin asymmetries - are performed with the \texttt{MontBlanc}~\cite{MontBlanc,MontBlancCodezenodo,AbdulKhalek:2022laj} and 
\texttt{Denali}~\cite{valerio_bertone_2024_10933177,valerio_bertone_2024_10933177_zenodo} frameworks,  
which enables efficient replica-by-replica evaluations and a flexible interface to FF-dependent SIDIS
observables, including the incorporation of EIC pseudodata. 
The strong coupling in this analysis is fixed to $\alpha_s(M_Z)=0.118$~\cite{ParticleDataGroup:2024cfk}, 
and the charm and bottom quark masses are set to $m_c=1.51$ GeV, and $m_b=4.92$ GeV~\cite{Bertone:2024taw}.

\section{Data sets}\label{data_sets}

In this section we summarize the experimental data sets included in our global QCD analysis.
We first describe the world polarized inclusive DIS and SIDIS measurements that define 
our {\tt pDIS+SIDIS base} fit, and we then introduce the projected EIC SIDIS pseudodata used to assess the
impact of future EIC measurements.

\subsection{Spin-dependent observables in polarized inclusive DIS and SIDIS}\label{subsec:world_data}

This analysis is based on a comprehensive set of measurements of spin-dependent observables in
polarized inclusive DIS and SIDIS off longitudinally polarized nucleon targets. 
For inclusive DIS, we consider data from the EMC~\cite{EuropeanMuon:1989yki}, SMC~\cite{SpinMuon:1998eqa},
and COMPASS~\cite{COMPASS:2015mhb,COMPASS:2016jwv} experiments at CERN; from the SLAC experiments
E142~\cite{E142:1996thl}, E143~\cite{E143:1998hbs}, E154~\cite{E154:1997xfa}, and E155~\cite{E155:2000qdr};
from HERMES at DESY~\cite{HERMES:1997hjr,HERMES:2006jyl}; and from Jefferson Lab measurements, including
Hall~A~\cite{JeffersonLabHallA:2016neg,Kramer:2002tt,JeffersonLabHallA:2004tea} and CLAS~\cite{CLAS:2014qtg}. 
These experiments provide measurements of the polarized structure function $g_1$, or of ratios such as
$g_1/F_1$ reconstructed from longitudinal double-spin asymmetries (see, \textit{e.g.}, 
Sec.~2.1 of the \texttt{NNPDFpol1.0} paper~\cite{Ball:2013lla}, for details).
The target in the aforementioned polarized DIS data sets is a proton, deuteron, or neutron.
We note that the present baseline dataset does not include the Jefferson Lab EG1b~\cite{CLAS:2015otq,CLAS:2017qga} and SANE~\cite{SANE:2018pwx} measurements. 
Their omission reflects the deliberately restricted DIS+SIDIS input adopted in this work and 
is consistent with our leading-twist NLO setup, in which TMCs and HT corrections are not included explicitly. 
A dedicated study including these datasets together with a more complete treatment of low-$W^2$ effects is left for future work.

For polarized SIDIS, we include charge-separated hadron production data from COMPASS~\cite{COMPASS:2010hwr} and
HERMES~\cite{HERMES:2018awh}.
Both experiments deliver measurements of the polarized semi-inclusive structure function $g_1^{h}$
normalized to its unpolarized counterpart $F_1^{h}$, for identified hadrons
$h=\pi^{+}, \pi^{-}, K^{+}, K^{-}$ and for proton and deuteron targets.
These SIDIS measurements provide enhanced sensitivity to the flavor decomposition of polarized PDFs,
and in particular help constrain the polarized sea-quark distributions
$\Delta \bar u$, $\Delta \bar d$, and $\Delta s$ through their dependence on FFs. 

The kinematic coverage of the fitted data spans approximately
$0.005 \lesssim x \lesssim 0.5$ and $1 \lesssim Q^2 \lesssim 100~\mathrm{GeV}^2$ for the fixed-target measurements,
with additional coverage at larger $x$ from Jefferson Lab. 
To ensure the validity of the leading-twist perturbative framework, we impose kinematic cuts on the photon
virtuality and on the invariant mass of the hadronic final state,
%
\begin{equation}
Q^2 \ge Q^2_{\rm cut}\,, \qquad
W^2 \equiv Q^2(\frac{1-x}{x}) \ge W^2_{\rm cut}\,,
\end{equation}
%
with the default choice $Q^2_{\rm cut}=1~\mathrm{GeV}^2$ and $W^2_{\rm cut}=4~\mathrm{GeV}^2$~\cite{Sato:2016tuz}. 
The cut on $Q^2$ suppresses regions where the strong coupling becomes large and perturbative predictions
are less reliable, while the cut on $W^2$ reduces sensitivity to power-suppressed effects beyond leading twist. 
We note that this $W^2$ cut is looser than the default choice adopted in some recent analyses
(e.g.\ $W^2_{\rm cut}=6.25~\mathrm{GeV}^2$ in Ref.~\cite{Bertone:2024taw}),
and therefore retains additional data points at lower $W^2$ that would be excluded by a more restrictive cut. 
Ref.~\cite{Sato:2016tuz} summarized the theoretical assumptions implied by this choice, 
and in Sec.~\ref{Results} we assess the robustness of the fit quality under these kinematic selections. 

Statistical and systematic uncertainties are provided separately for most measurements, while detailed
information on correlated systematics is available only for a subset of data sets. 
In line with the common treatment adopted in global QCD analyses, whenever an experimental covariance matrix
or correlated normalization uncertainties are provided (notably for selected polarized DIS and polarized SIDIS data sets),
we incorporate the corresponding correlation information.
For measurements where no correlation information is available, systematic and statistical uncertainties are
added in quadrature~\cite{Bertone:2024taw}).

\subsection{EIC SIDIS pseudodata}\label{subsec:eic_pseudodata}

To enhance the flavor separation of sea polarized PDFs and extend the kinematic reach
towards small $x$, we incorporate projected EIC measurements in polarized SIDIS.
We adopt the EIC pseudodata strategy developed within the ECCE studies
of Ref.~\cite{Adkins:2022jfp} and, in particular, we follow the detector-level framework
presented in Ref.~\cite{VanHulse:2023uga}. 
We consider charge-separated hadron production
$\pi^+$, $\pi^-$, $K^+$  and $K^-$ in longitudinally polarized electron-proton scattering for two beam-energy 
configurations, $E_e \times E_p = 5 \times 41~\mathrm{GeV^2}$ and $18 \times 275~\mathrm{GeV^2}$,
which provide complementary sensitivity across $(x,Q^2)$ and extend the SIDIS reach down to
very small $x$.

In the ECCE simulation setup of Ref.~\cite{VanHulse:2023uga}, events are selected in the DIS regime with
$Q^2>1~\mathrm{GeV}^2$, a photon-nucleon invariant mass cut $W^2>10~\mathrm{GeV}^2$ to suppress
the resonance region, and an inelasticity selection $0.01 < y < 0.95$ to control radiative effects
and ensure robust reconstruction. 
Identified hadrons are required to satisfy $z > 0.01$ (to limit target-fragmentation contributions), 
and no explicit rapidity requirement is imposed. Statistical uncertainties are scaled to an
integrated luminosity of $10~\mathrm{fb}^{-1}$, assuming realistic longitudinal polarization of
$70\%$ for both beams; in addition, an overall $\sim 2\%$ polarization (scale) uncertainty is quoted
as a separate normalization component~\cite{VanHulse:2023uga}. 
The kinematic binning in $(x_B,Q^2,z)$ used for the projections follows Ref.~\cite{VanHulse:2023uga}. 

For the purposes of this global analysis, we construct EIC pseudodata for the longitudinal
double-spin asymmetries $A_1^h(x,z,Q^2)$ by starting from NLO theory predictions computed
with the \texttt{MontBlanc}~\cite{MontBlanc,MontBlancCodezenodo,AbdulKhalek:2022laj} and 
\texttt{Denali}~\cite{valerio_bertone_2024_10933177,valerio_bertone_2024_10933177_zenodo} frameworks 
in the same SIDIS kinematics, and then applying a smearing procedure consistent with the projected uncertainties. 
In constructing these central NLO predictions, we use the helicity PDFs from our {\tt pDIS+SIDIS base} fit, 
together with the \texttt{NNPDF4.0} NLO unpolarized PDFs~\cite{NNPDF:2021njg} (with perturbative charm) and the \texttt{MAPFF1.0} NLO FFs for charged pions and kaons~\cite{Khalek:2021gxf,AbdulKhalek:2022laj}.
Concretely, for each kinematic bin $i$ we take the NLO prediction $A^{h,\mathrm{th}}_{1,i}$ and define a
pseudo-measurement as

\begin{equation}
A^{h,\mathrm{(EIC)}}_{1,i} \;=\; A^{h,\mathrm{th}}_{1,i} + r_i\,\sigma^{\mathrm{unc}}_{i}
\;+\; \delta_{\mathrm{pol}}\,A^{h,\mathrm{th}}_{1,i},
\end{equation}

where $r_i$ is a standard normal variate modeling an uncorrelated bin-by-bin fluctuation, and
$\sigma^{\mathrm{unc}}_{i}$ is the absolute projected uncertainty for that bin (constructed from the
relative uncertainty quoted by the ECCE projections). 
The parameter $\delta_{\mathrm{pol}}$ is a correlated systematic shift with width $2\%$, 
applied coherently to all bins to emulate the overall beam-polarization
(scale) uncertainty discussed in Ref.~\cite{VanHulse:2023uga}. 
In our numerical implementation, we additionally 
enforce sign stability of the smeared asymmetries in each bin by resampling the uncorrelated fluctuation if a
sign flip occurs; this avoids unphysical pseudo-measurements in bins where $A_1^h$ is small compared to its
uncertainty.

\begin{figure*}[htb]
\vspace{0.50cm}
\centering
\includegraphics[width=0.49\textwidth]{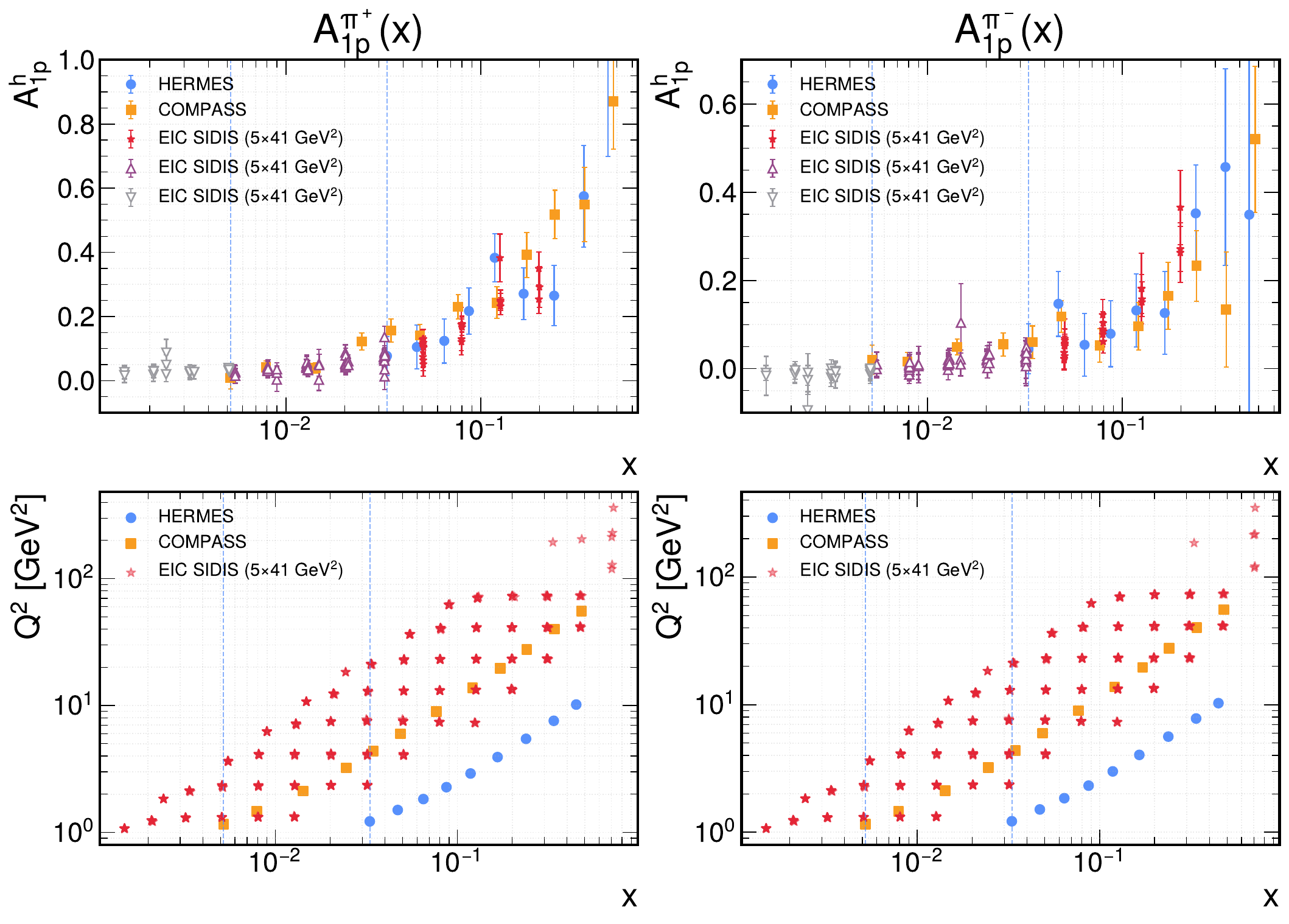}
\includegraphics[width=0.49\textwidth]{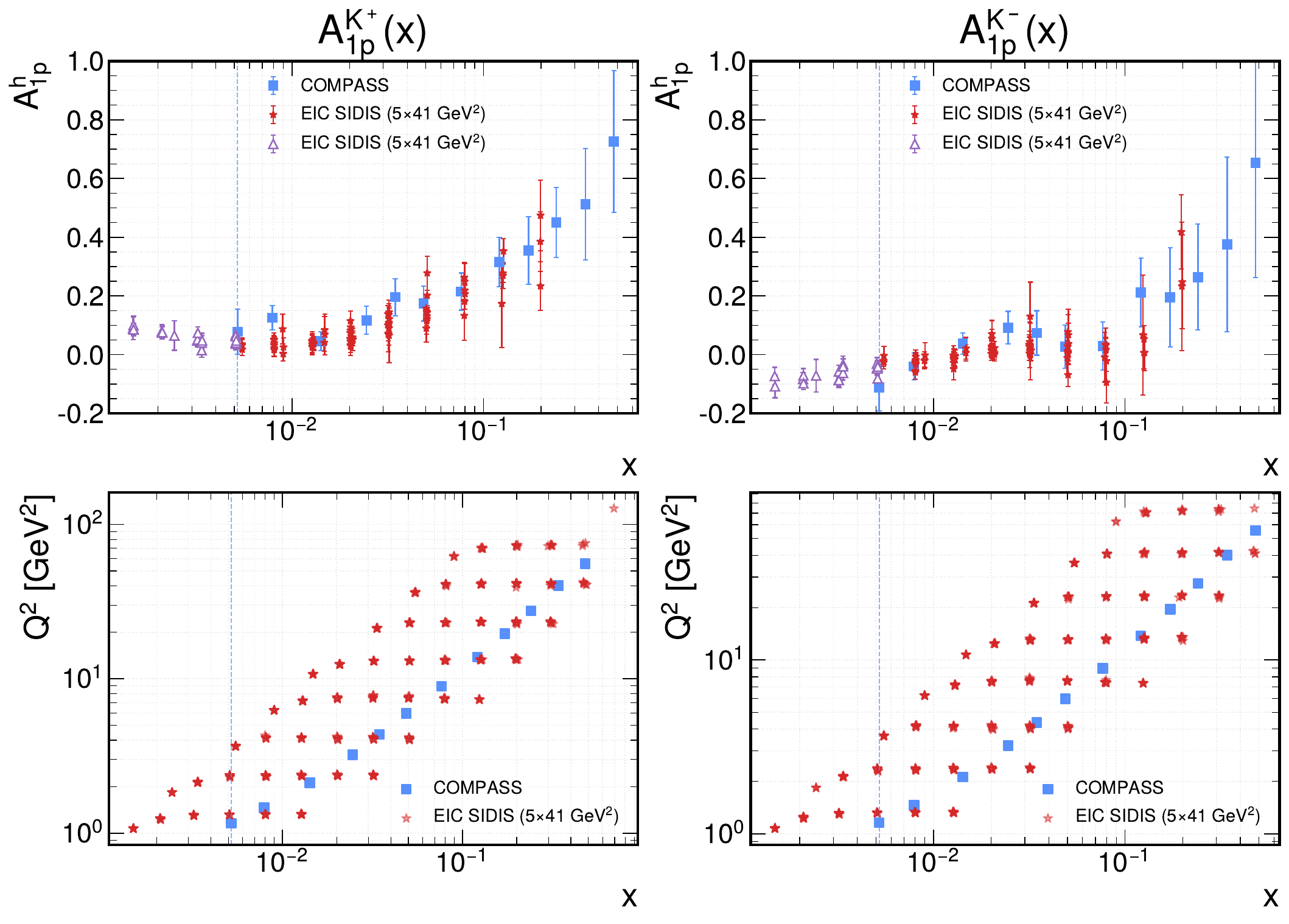}
\begin{center}
\caption{\small
Comparison of representative proton SIDIS asymmetries and kinematic coverage for charged-pion (left) and charged-kaon (right) production. 
In each case, the upper panels display representative $A_{1p}^{h}(x)$ comparisons in a restricted EIC kinematic slice,
$0.30<z<0.60$ and $1.0<Q^2<20.0~\mathrm{GeV}^2$,
chosen to provide a clear observable-level comparison in a region relevant to the existing fixed-target measurements.
The lower panels show the corresponding broader EIC coverage for
$0.10<z<0.90$ and $Q^2>1.0~\mathrm{GeV}^2$.}
\label{fig:EIC_HERMES_COMPASS}
\end{center}
\end{figure*}

In Fig.~\ref{fig:EIC_HERMES_COMPASS}, we illustrate the impact of the projected EIC SIDIS ($5 \times 41~\mathrm{GeV^2}$) 
measurements for charged pions and kaons. 
The upper panels show representative $A_{1p}^{h}(x)$ comparisons with the existing fixed-target data,
where the EIC points are restricted to the slice $0.30<z<0.60$ and $1.0<Q^2<20.0~\mathrm{GeV}^2$ in order to provide a readable observable-level comparison.
The lower panels instead show the broader EIC coverage for $0.10<z<0.90$ and $Q^2>1.0~\mathrm{GeV}^2$,
thereby emphasizing the substantial extension of the SIDIS kinematic reach toward lower $x$ and over a wider $Q^2$ range.
The figure illustrates both the continuity of the projected EIC pseudodata with the existing fixed-target measurements in the overlap region and, more importantly, the substantial extension of the accessible SIDIS phase space toward lower $x$ and over a wider $Q^2$ lever arm.

%
%
%
%

The projected EIC SIDIS asymmetries provide strong new constraints on polarized PDFs down to $x \sim 10^{-5}$.
At moderate and large $x$, they sharpen the flavor separation through hadron tagging and its sensitivity to
FFs, while at small $x$ they significantly improve the reach and precision of the extracted
sea-quark helicity distributions and, indirectly, the gluon helicity distribution through the global fit interplay.

\begin{figure*}[htb]
\vspace{0.50cm}
\centering
\includegraphics[width=0.70\textwidth]{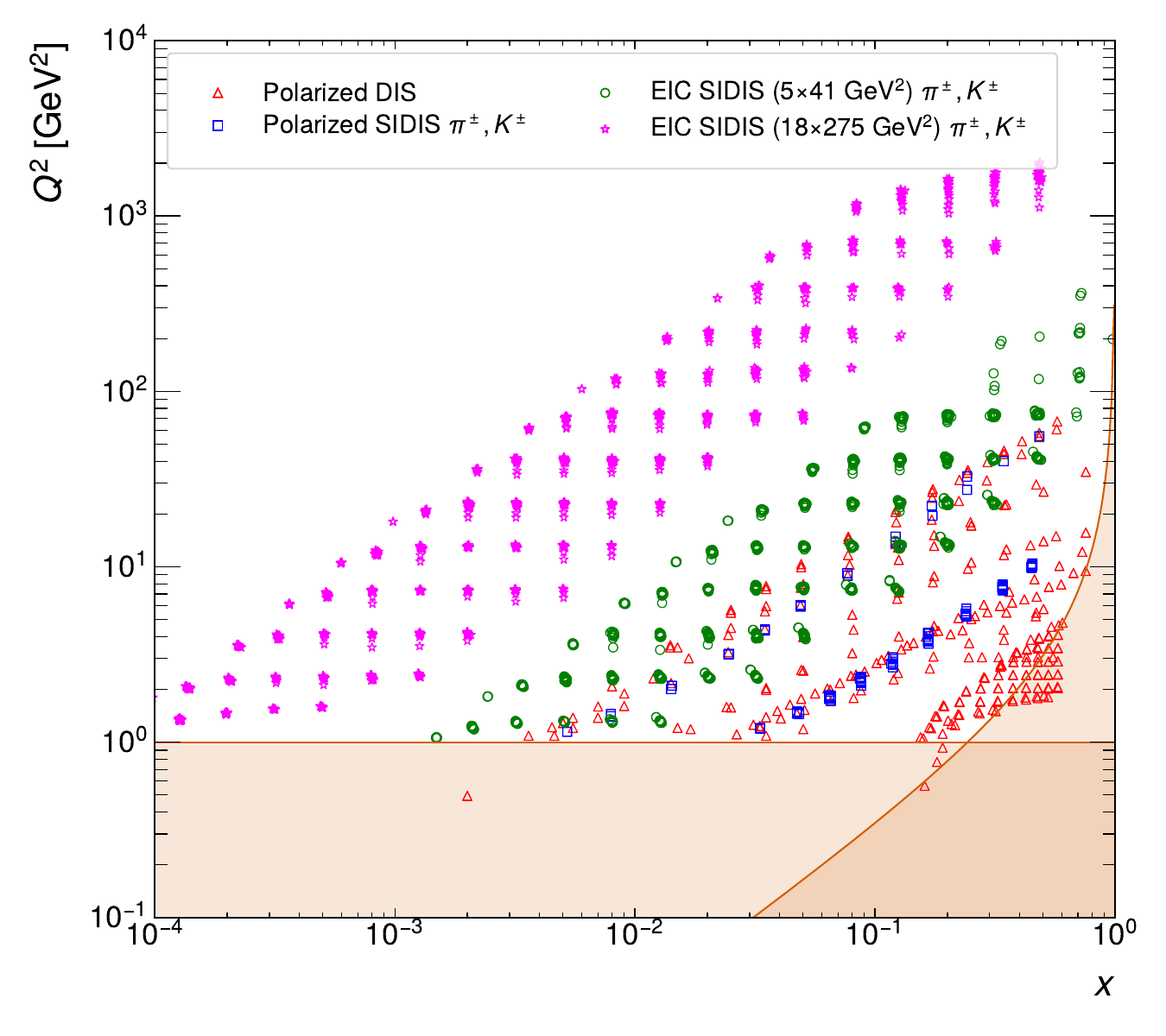}
\begin{center}
\caption{ \small 
Kinematic coverage in the $(x,Q^{2})$ plane for the datasets included
in this analysis: polarized inclusive DIS (triangles) and polarized SIDIS
$\pi^+$, $\pi^-$, $K^+$  and $K^-$ from HERMES and COMPASS (squares), together with projected
EIC SIDIS asymmetries for $\pi^+$, $\pi^-$, $K^+$  and $K^-$ at $5\times41~\mathrm{GeV^2}$ (circles) and
$18\times275~\mathrm{GeV^2}$ (stars). 
The shaded region corresponds to the excluded domain
imposed by the cuts $Q^{2} > 1~\mathrm{GeV}^{2}$ and $W^{2} > 4~\mathrm{GeV}^{2}$;
the horizontal line indicates the $Q^{2}$ threshold, and the curved line
indicates the $W^{2}$ boundary.}
\label{fig:xq2_coverage}
\end{center}
\end{figure*}

Figure~\ref{fig:xq2_coverage} shows the kinematic coverage in the $(x,Q^{2})$ plane
for the polarized inclusive DIS data, the polarized SIDIS measurements from HERMES
and COMPASS, and the projected EIC SIDIS asymmetries for identified
$\pi^+$, $\pi^-$, $K^+$  and $K^-$ production at the $E_e\times E_p = 5\times41~\mathrm{GeV^2}$ and $E_e\times E_p = 18\times275~\mathrm{GeV^2}$
beam-energy configurations. 
As discussed above, a common set of DIS cuts is imposed throughout the analysis,
$Q^{2} > 1~\mathrm{GeV}^{2}$ and $W^{2} > 4~\mathrm{GeV}^{2}$, with the excluded
regions indicated by the shaded area (the $Q^{2}$ cut) and the curved boundary
(the $W^{2}$ cut). 
As can be seen from Fig.~\ref{fig:xq2_coverage}, while the existing polarized 
DIS and SIDIS data primarily constrain the 
moderate-to-large $x$ region at relatively low-to-intermediate $Q^{2}$, the EIC
projections provide a substantially extended lever arm in both $x$ and $Q^{2}$,
including dense coverage at fixed $x$ over multiple $Q^{2}$ values. 
This enhanced lever arm is particularly valuable for improving sensitivity to the gluon and sea-quark polarized 
distributions through QCD evolution. 

%
%
%
%

We perform three QCD fits in order to quantify the impact of EIC projected measurements. 
The baseline determination, denoted as the {\tt pDIS+SIDIS base} fit, includes the available 
polarized DIS and SIDIS data sets described in Sec.~\ref{data_sets}, namely inclusive polarized DIS measurements from
EMC, SMC, the SLAC experiments E142, E143, E154, and E155, HERMES, COMPASS, and Jefferson Lab (Hall~A and CLAS),
together with charge-separated SIDIS $\pi^{\pm}$ and $K^{\pm}$ measurements from HERMES and COMPASS. 
Starting from this baseline, we then include the EIC projected SIDIS pseudodata at
$E_e\times E_p = 5\times41~\mathrm{GeV^2}$   and at $18\times275~\mathrm{GeV^2}$, producing the two additional fits
{\tt pDIS+SIDIS+EIC(5$\times$41)} and {\tt pDIS+SIDIS+EIC(18$\times$275)}.
We treat the two EIC energy configurations separately, which allows us to disentangle their individual impact, since the two setups probe 
complementary kinematic regions. 
The lower-energy configuration provides an important benchmark for the first-stage EIC sensitivity, 
while the higher-energy configuration primarily enlarges the $(x,Q^2)$ lever arm and extends the reach further 
into the small-$x$ region.

A detailed summary of the experimental data sets entering the {\tt pDIS+SIDIS base} fit,
together with the corresponding number of data points $N_{\rm dat}$ and their contribution
to the goodness-of-fit $\chi^2/N_{\rm dat}$ evaluated at the best-fit point, is provided in
Table~\ref{tab:chi2-summary}. 

%
%
\begin{table}[t]
\centering
\small
\caption{\small
Summary of the polarized DIS and SIDIS data sets included in the {\tt pDIS+SIDIS base} fit.
For each experiment and observable we report the number of data points, $N_{\rm dat}$,
and the corresponding contribution to the goodness-of-fit, $\chi^2/N_{\rm dat}$,
evaluated at the best-fit point. }
\label{tab:chi2-summary}
\begin{tabular*}{\linewidth}{@{\extracolsep{\fill}} l l l r r}
\toprule
Experiment & Ref. & Observable & $N_{\rm dat}$ & $\chi^2/N_{\rm dat}$ \\
\midrule
COMPASS & \cite{COMPASS:2010hwr} & $A_1^{p}(K^-)$         & 12 & 0.727 \\
COMPASS & \cite{COMPASS:2010hwr} & $A_1^{p}(K^+)$         & 12 & 0.719 \\
COMPASS & \cite{COMPASS:2010hwr} & $A_1^{p}(\pi^-)$       & 12 & 1.201 \\
COMPASS & \cite{COMPASS:2010hwr} & $A_1^{p}(\pi^+)$       & 12 & 2.500 \\
COMPASS & \cite{COMPASS:2010hwr} & $A_1^{d}(K^-)$         & 10 & 0.891 \\
COMPASS & \cite{COMPASS:2010hwr} & $A_1^{d}(K^+)$         & 10 & 0.321 \\
COMPASS & \cite{COMPASS:2010hwr} & $A_1^{d}(\pi^-)$       & 10 & 0.448 \\
COMPASS & \cite{COMPASS:2010hwr} & $A_1^{d}(\pi^+)$       & 10 & 0.320 \\
HERMES  & \cite{HERMES:2018awh}  & $A_1^{d}(K^-)$            &  9 & 0.612 \\
HERMES  & \cite{HERMES:2018awh}  & $A_1^{d}(K^+)$            &  9 & 1.459 \\
HERMES  & \cite{HERMES:2018awh}  & $A_1^{d}(\pi^-)$          &  9 & 1.265 \\
HERMES  & \cite{HERMES:2018awh}  & $A_1^{d}(\pi^+)$          &  9 & 0.334 \\
HERMES  & \cite{HERMES:2018awh}  & $A_1^{p}(\pi^-)$          &  9 & 1.103 \\
HERMES  & \cite{HERMES:2018awh}  & $A_1^{p}(\pi^+)$          &  9 & 1.889 \\
\midrule
E142    & \cite{E142:1996thl}    & $g_1^{n}$                 & 8  & 0.714 \\
E143    & \cite{E143:1998hbs}    & $g_1^{d}$                 & 27 & 1.315 \\
E143    & \cite{E143:1998hbs}    & $g_1^{p}$                 & 27 & 0.902 \\
E154    & \cite{E154:1997xfa}    & $g_1^{n}$                 & 11 & 0.324 \\
E155    & \cite{E155:2000qdr}    & $g_1^{p}/F_1^{p}$         & 24 & 0.630 \\
E155    & \cite{E155:2000qdr}    & $g_1^{n}/F_1^{n}$         & 24 & 0.644 \\
EMC     & \cite{EuropeanMuon:1989yki} & $g_1^{p}$            & 10 & 0.639 \\
JLab E06-014 & \cite{JeffersonLabHallA:2016neg} & $g_1^{n}/F_1^{n}$ & 4 & 2.010 \\

JLab E97-103 & \cite{Kramer:2002tt}              & $g_1^{n}$         & 2 & 0.414 \\
JLab E99-117 & \cite{JeffersonLabHallA:2004tea}              & $g_1^{n}/F_1^{n}$  & 1 & 0.006 \\

JLab EG1-DVCS & \cite{CLAS:2014qtg}             & $g_1^{d}/F_1^{d}$  & 19 & 0.162 \\
JLab EG1-DVCS & \cite{CLAS:2014qtg}             & $g_1^{p}/F_1^{p}$  & 21 & 0.132 \\
SMC     & \cite{SpinMuon:1998eqa}  & $g_1^{d}$                     & 12 & 1.315 \\
SMC     & \cite{SpinMuon:1998eqa}  & $g_1^{p}$                     & 12 & 0.373 \\
COMPASS & \cite{COMPASS:2016jwv} & $g_1^{d}$                       & 15 & 0.331 \\
COMPASS & \cite{COMPASS:2015mhb} & $g_1^{p}$                       & 17 & 0.455 \\
HERMES  & \cite{HERMES:1997hjr} & $g_1^{n}$                        & 9  & 0.226 \\
HERMES  & \cite{HERMES:2006jyl} & $g_1^{d}$                        & 15 & 1.197 \\
HERMES  & \cite{HERMES:2006jyl} & $g_1^{p}$                        & 15 & 1.005 \\
\midrule
\multicolumn{3}{l}{Total} & 415 &  0.727 \\
\bottomrule
\end{tabular*}
\end{table}
%
%


\subsection{Fitting methodology}\label{Fitting_methodology}

The helicity-dependent PDFs are extracted using a Monte Carlo (MC) replica methodology combined with a 
neural-network (NN) parameterization, closely following the strategy adopted in {\tt MAPPDFpol1.0} determinations~\cite{Bertone:2024taw}. 
Experimental uncertainties (including those of the EIC pseudodata) are propagated to the fitted polarized PDFs 
by generating an ensemble of statistically equivalent replicas of the input dataset. 
Whenever an experimental covariance matrix or correlated normalization uncertainties are available, 
the corresponding correlations are included; otherwise statistical and systematic uncertainties are added in quadrature, 
in line with the treatment described in Sec.~\ref{data_sets}. 

For each replica $k$, a fluctuated data vector may be written schematically as
%
\begin{equation}
  y^{(k)} = y + L\,r^{(k)}\,,\qquad
  r^{(k)} \sim \mathcal{N}(0,I)\,,
\end{equation}
%
where $C=LL^{T}$ is the total covariance matrix.

\paragraph{Parametrization and flavor basis:}

At the input scale $Q_0=1~\mathrm{GeV}$, polarized PDFs are parameterized by a single 
feed-forward NN implemented in the \texttt{MontBlanc} framework~\cite{MontBlanc,AbdulKhalek:2022laj}.
To construct the independent PDF basis
%
\begin{equation}
\left\{
\Delta f_u,\ \Delta f_{\bar u},\ \Delta f_d,\ \Delta f_{\bar d},\ \Delta f_s,\ \Delta f_{\bar s},\ \Delta f_g
\right\},
\end{equation}
%
in the way explained in the next subsection, i.e.\ allowing $\Delta f_s$ and $\Delta f_{\bar s}$ to be fitted independently, the run configuration adopts one input node ($x$), a single hidden layer with 10 nodes, and 7 output nodes,

%
%
\begin{equation}
\begin{tikzpicture}[baseline=(current bounding box.center),
    innode/.style={circle, draw, minimum size=0.75cm, inner sep=1pt},
    node/.style={circle, draw, minimum size=0.5cm, inner sep=1pt},
    outnode/.style={circle, draw, minimum size=0.75cm, inner sep=1pt},
    >=stealth]
    \node[innode] (i1) at (0, -2.8) {$x$};
    \foreach \j in {1,...,10}
    \node[node] (h\j) at (2.5, -\j*0.55) {};
    \node[outnode, label=right:{\small $\Delta f_u$}] (o1) at (5.2, -0.5) {};
    \node[outnode, label=right:{\small $\Delta f_{\bar u}$}] (o2) at (5.2, -1.3) {};
    \node[outnode, label=right:{\small $\Delta f_d$}] (o3) at (5.2, -2.1) {};
    \node[outnode, label=right:{\small $\Delta f_{\bar d}$}] (o4) at (5.2, -2.9) {};
    \node[outnode, label=right:{\small $\Delta f_s$}] (o5) at (5.2, -3.7) {};
    \node[outnode, label=right:{\small $\Delta f_{\bar s}$}] (o6) at (5.2, -4.5) {};
    \node[outnode, label=right:{\small $\Delta f_g$}] (o7) at (5.2, -5.3) {};
    \foreach \j in {1,...,10}
        \draw[->, opacity=0.3] (i1) -- (h\j);
    \foreach \j in {1,...,10}
        \foreach \k in {1,...,7}
            \draw[->, opacity=0.3] (h\j) -- (o\k);
    \node at (0, 0.2) {\small Input};
    \node at (2.5, 0.2) {\small Hidden};
    \node at (5.2, 0.2) {\small Output};
    \node at (2.6, 0.55) {$[1 \qquad\qquad\qquad\ \ \  10 \qquad\qquad\qquad\qquad 7]$};
\end{tikzpicture}
\end{equation}
%
%
%

\paragraph{Positivity constraints:}

The fit enforces the standard LO-motivated positivity bound~\cite{Bertone:2024taw,Kataev:2003jv},
%
\begin{equation}
  |\Delta f_i(x,Q^2)| \le f_i(x,Q^2)\,,
\end{equation}
%
using an unpolarized PDF ensemble to provide the $x$-dependent upper limit.
In the default configuration, positivity is implemented by construction at the parameterization scale through
%
\begin{equation}
\begin{aligned}
  \Delta f_i^{(k)}(x,Q_0^2)
  &=
  \left[2\,\mathrm{NN}_i(x)-1\right]\,
  f_i^{(r_k)}(x,Q_0^2)\,,
  \\
  &\qquad i=g,u,\bar{u},d,\bar{d},s,\bar{s}\,.
\end{aligned}
\label{eq:posnet_hamzeh}
\end{equation}
%
where $f_i^{(r_k)}$ denotes a replica of the chosen unpolarized PDF set.
With the replica policy used here, the unpolarized-PDF replica index $r_k$ is selected randomly 
for each fitted polarized replica, thereby propagating the uncertainty of the positivity bound into the polarized PDFs.
An alternative (looser) option, available in the framework, is to impose the positivity bound 
using the unpolarized central set inflated by a multiplicative factor; this option is disabled in the present analysis.

\paragraph{Theory settings and external inputs:}

All theory predictions entering the fit are computed consistently at NLO in QCD. 
Polarized PDFs are evolved from $Q_0$ using NLO DGLAP evolution in the $\overline{\mathrm{MS}}$ scheme.
The strong coupling is fixed to $\alpha_s(M_Z)=0.118$ and the heavy-flavor thresholds are 
set to $m_c=1.51~\mathrm{GeV}$ and $m_b=4.92~\mathrm{GeV}$~\cite{Bertone:2024taw}. 
Unpolarized PDFs required for the positivity bounds and for unpolarized denominators are taken 
from the \texttt{NNPDF4.0} NLO set with perturbative charm~\cite{NNPDF:2021njg}. 
For SIDIS observables, fragmentation functions are taken from the \texttt{MAPFF1.0} NLO sets 
for charged pions and kaons~\cite{Khalek:2021gxf,AbdulKhalek:2022laj}. 
For each Monte Carlo replica of the fit, replicas of the external unpolarized PDF and FF sets are
selected accordingly, and hence, that the corresponding
uncertainties are consistently propagated to the theory predictions and, ultimately, to the fitted
polarized PDFs.  
Since \texttt{NNPDF4.0} and \texttt{MAPFF1.0} are themselves neural-network Monte Carlo
determinations, this procedure preserves methodological consistency across inputs and ensures a
faithful propagation of external-set uncertainties in the QCD fit.

\paragraph{Optimization, cross-validation, and replica selection:}

For each data replica $k$, the NN parameters are determined by minimizing the (correlated) $\chi^2$ function
%
\begin{equation}
  \chi^2_k =
  \left(\bm{T}[\Delta f^{(k)}]-\bm{y}^{(k)}\right)^{T}
  \bm{C}^{-1}
  \left(\bm{T}[\Delta f^{(k)}]-\bm{y}^{(k)}\right),
\end{equation}
%
where $\bm{T}$ denotes the vector of theory predictions for all fitted points.
The minimization is performed with the Levenberg-Marquardt algorithm as 
implemented in \textsc{Ceres-Solver}~\cite{Agarwal_Ceres_Solver_2022}, 
with a maximum number of iterations.  
To mitigate overfitting, cross-validation is applied using an 80\%/20\% training/validation split 
for datasets with more than 10 points, while smaller datasets are used fully for training. 
Finally, replicas that fail to converge or yield an anomalously poor description of the data are 
discarded according to the configured $\chi^2$ tolerance($\chi^2_{\mathrm{Max}}=3.0$). 
The resulting ensemble $\{\Delta f^{(k)}(x,Q^2)\}_{k=1}^{N_{\mathrm{rep}}}$ provides a Monte Carlo 
representation of the probability distribution in functional space, from which central values and 
uncertainties are computed as replica statistics.  

All steps of the analysis - PDF evolution, computation of polarized inclusive DIS and polarized SIDIS observables, 
replica generation, and minimization - are performed 
within the \texttt{MontBlanc}~\cite{MontBlanc,MontBlancCodezenodo,AbdulKhalek:2022laj} and 
\texttt{Denali}~\cite{valerio_bertone_2024_10933177,valerio_bertone_2024_10933177_zenodo} fitting frameworks.

\section{Results}\label{Results}

In this section we present the results of our NLO global determination of
helicity-dependent PDFs and quantify the impact of the projected EIC SIDIS
measurements. We consider three fit scenarios throughout:
(i) the {\tt pDIS+SIDIS base} fit to existing world data (world polarized DIS and SIDIS data only);
(ii) {\tt pDIS+SIDIS+EIC(5$\times41~\mathrm{GeV^2}$)} including EIC SIDIS pseudodata at
$E_e\times E_p=5\times41~\mathrm{GeV^2}$; and
(iii) {\tt pDIS+SIDIS+EIC(18$\times275~\mathrm{GeV^2}$)} including EIC SIDIS pseudodata at
$E_e\times E_p=18\times275~\mathrm{GeV^2}$.  
The present analysis should therefore be interpreted as a {\tt DIS+SIDIS baseline} to {\tt EIC-SIDIS} impact study,  
and the inclusion of polarized RHIC proton-proton data, which provide important complementary constraints 
on $\Delta g$ and on sea-quark flavor separation, is deferred to future work.

\subsection{Fit quality}\label{subsec:Fit-quality}

We quantify the fit quality by evaluating the $\chi^2$ per data point for each fitted data set
at the best-fit point of the corresponding Monte Carlo determination.
A detailed breakdown of the polarized DIS and SIDIS measurements entering the
{\tt pDIS+SIDIS base} fit is provided in Table~\ref{tab:chi2-summary}, where we report the number of
data points $N_{\rm dat}$ and the corresponding $\chi^2/N_{\rm dat}$ contribution for each experiment
and observable. 
The total fit quality for the baseline determination is $\chi^2/N_{\rm dat}=0.727$ for $N_{\rm dat}=415$. 
At the level of individual subsets, the description is generally very good: 
the majority of subsets have $\chi^2/N_{\rm dat}<1$ (and many well below unity), while a small number exhibit larger values. 
The most pronounced tensions are observed for the COMPASS proton $\pi^+$ asymmetry and the 
JLab E06-014 subset (see Table~\ref{tab:chi2-summary}), with a few additional subsets showing mild 
tension at the level $\chi^2/N_{\rm dat}\sim 1.2$-$1.5$.
Experimental correlations are included whenever an experimental covariance matrix or correlated normalisation
uncertainties are provided; otherwise statistical and systematic uncertainties are added in quadrature.
Our baseline kinematic selections are $Q^2\ge 1~\mathrm{GeV}^2$ and $W^2\ge 4~\mathrm{GeV}^2$.
The cut on $Q^2$ suppresses regions where the strong coupling becomes large and fixed-order perturbative
predictions are less reliable, while the cut on $W^2$ reduces sensitivity to power-suppressed effects beyond
leading twist. This $W^2$ cut is looser than that adopted in some recent analyses (including the MAP study,
which uses a tighter $W^2$ cut and notes that some Jefferson Lab subsets are excluded by its kinematic
selections), and therefore retains additional points at lower $W^2$ that would be removed by more restrictive
choices.

\subsection{\texorpdfstring{$z_{\min}$}{zmin} dependence and $\chi^2$ stability of the global fit}\label{subsec:chi2_zscan}

\begin{figure*}[htb]
\vspace{0.50cm}
\centering
\includegraphics[width=0.48\textwidth]{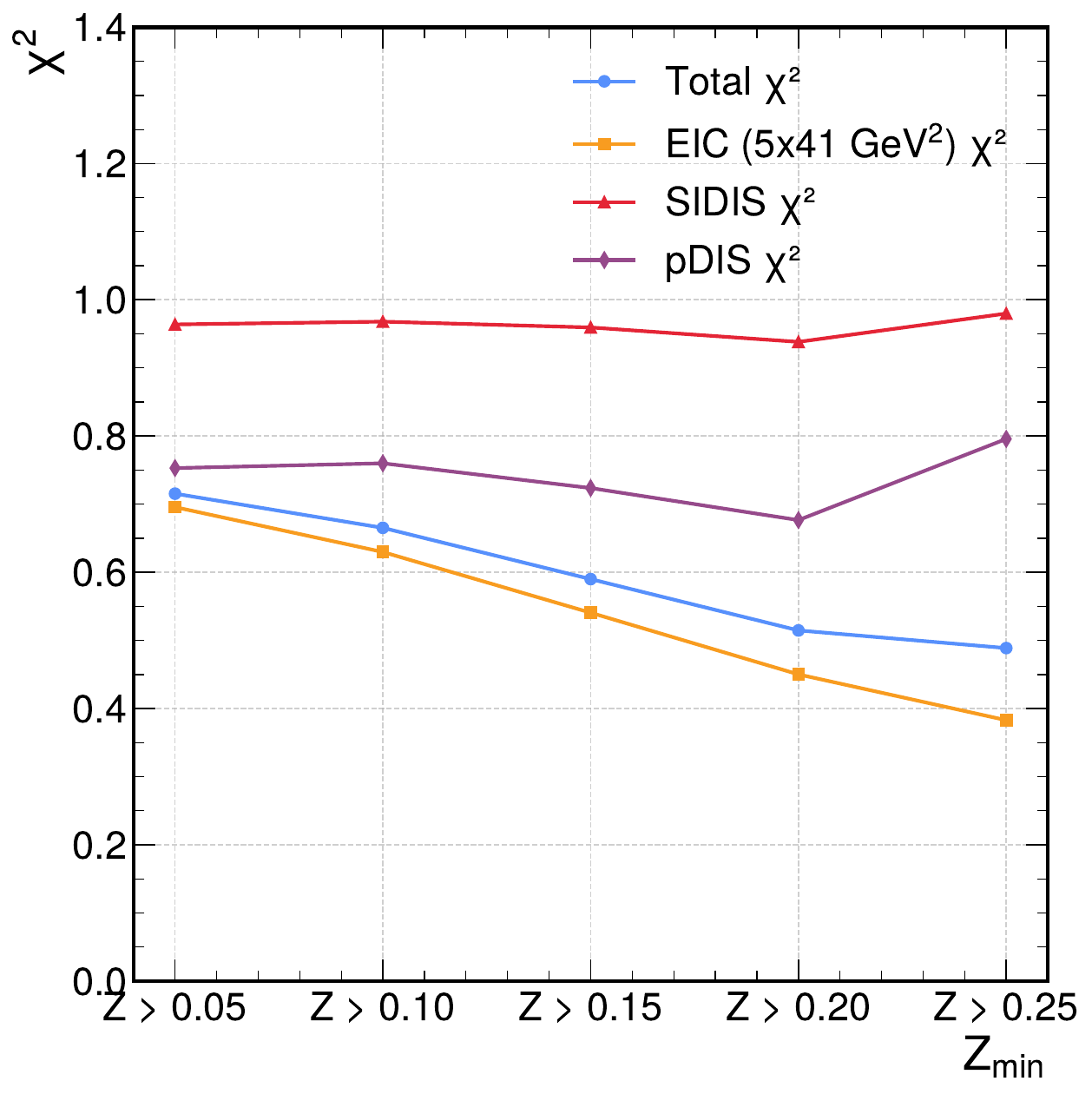}
\includegraphics[width=0.48\textwidth]{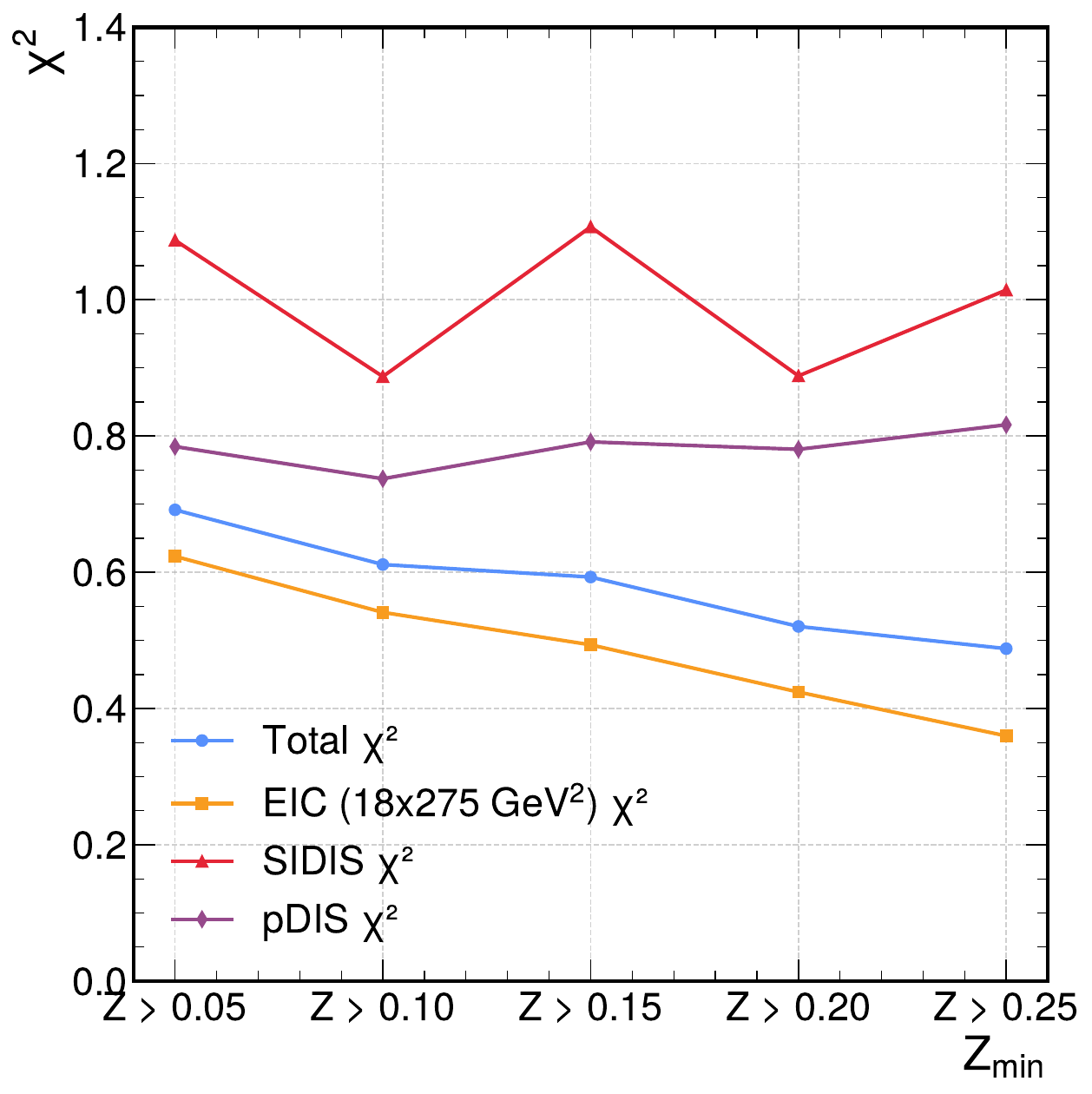}
\begin{center}
\caption{ \small 
Impact of different $z$-cut thresholds on the quality of the global fit.
The left (right) panel shows the $\chi^2$ values for the total dataset and its individual components,
polarized DIS (pDIS), SIDIS, and EIC pseudodata, as a function of the minimum $z$ cut,
separately for the $5\times41~\mathrm{GeV^2}$ and ($18\times275~\mathrm{GeV^2}$) configurations. }
\label{fig:chi2_vs_z}
\end{center}
\end{figure*}

In this section, we quantify the dependence on $z_{\min}$ by scanning the goodness-of-fit as 
a function of the minimum $z$ requirement for the EIC pseudodata. 
The corresponding $\chi^2$ profiles are shown in 
Fig.~\ref{fig:chi2_vs_z} for the two EIC energy configurations. 
For both $5 \times 41~\mathrm{GeV^2}$ and $18 \times 275~\mathrm{GeV^2}$, the total $\chi^2$ exhibits a smooth dependence on $z_{\min}$, 
with a mild preference for tighter $z$ cuts. Importantly, the partial contributions from polarized DIS and SIDIS
remain stable across the scan, indicating that the overall description of the existing world data is not
driven by the specific $z_{\min}$ choice adopted for the EIC pseudodata. Based on these trends, and to
provide a conservative baseline against potential low-$z$ hadronization effects, we take $z_{\min}>0.2$
as default, while also reporting results for $z_{\min}>0.1$ to illustrate the sensitivity of the conclusions
to this selection. 

%
%
%
%

\begin{figure*}[htb]
\vspace{0.50cm}
\centering
\includegraphics[width=0.90\textwidth]{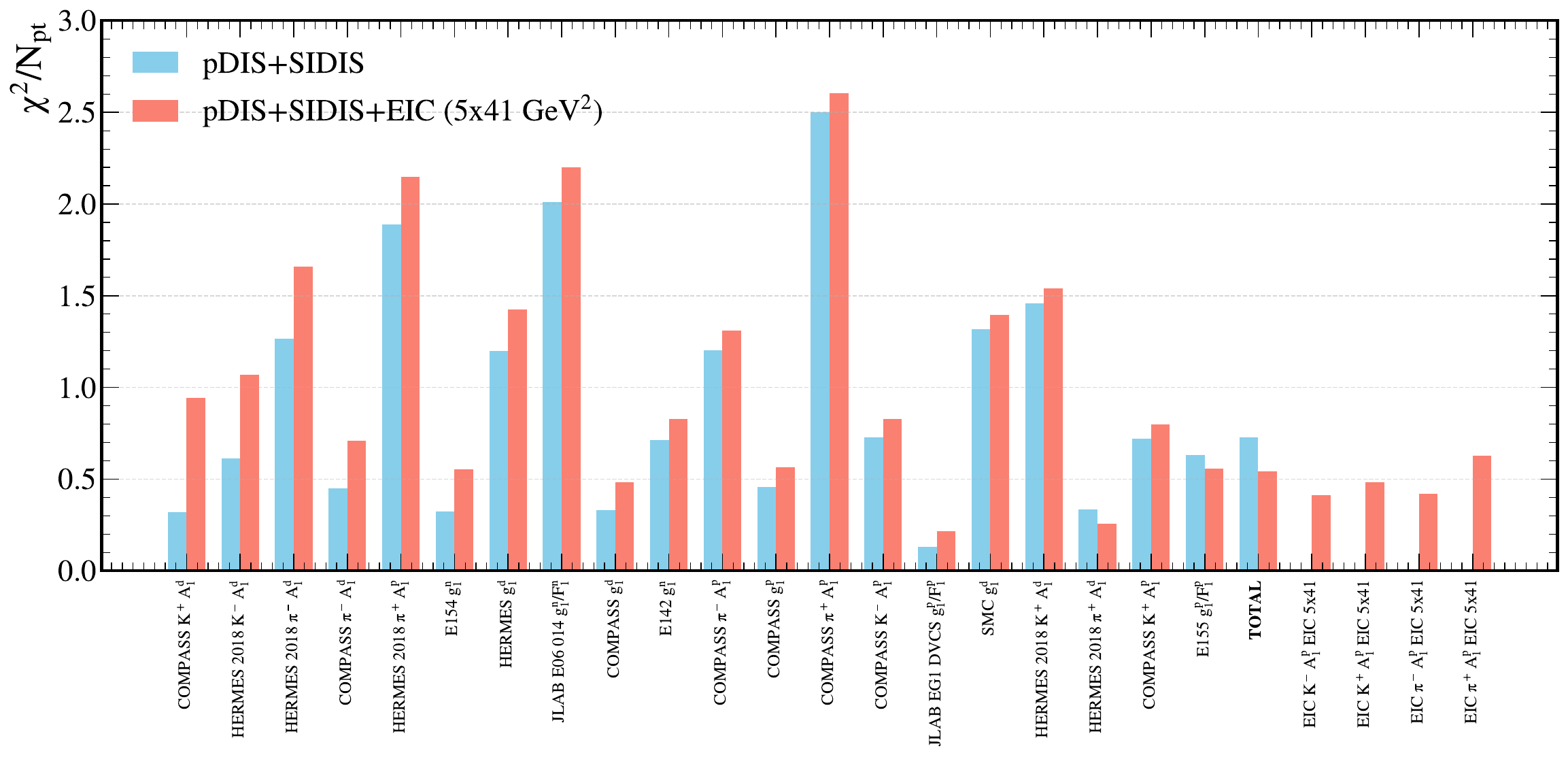}
\includegraphics[width=0.90\textwidth]{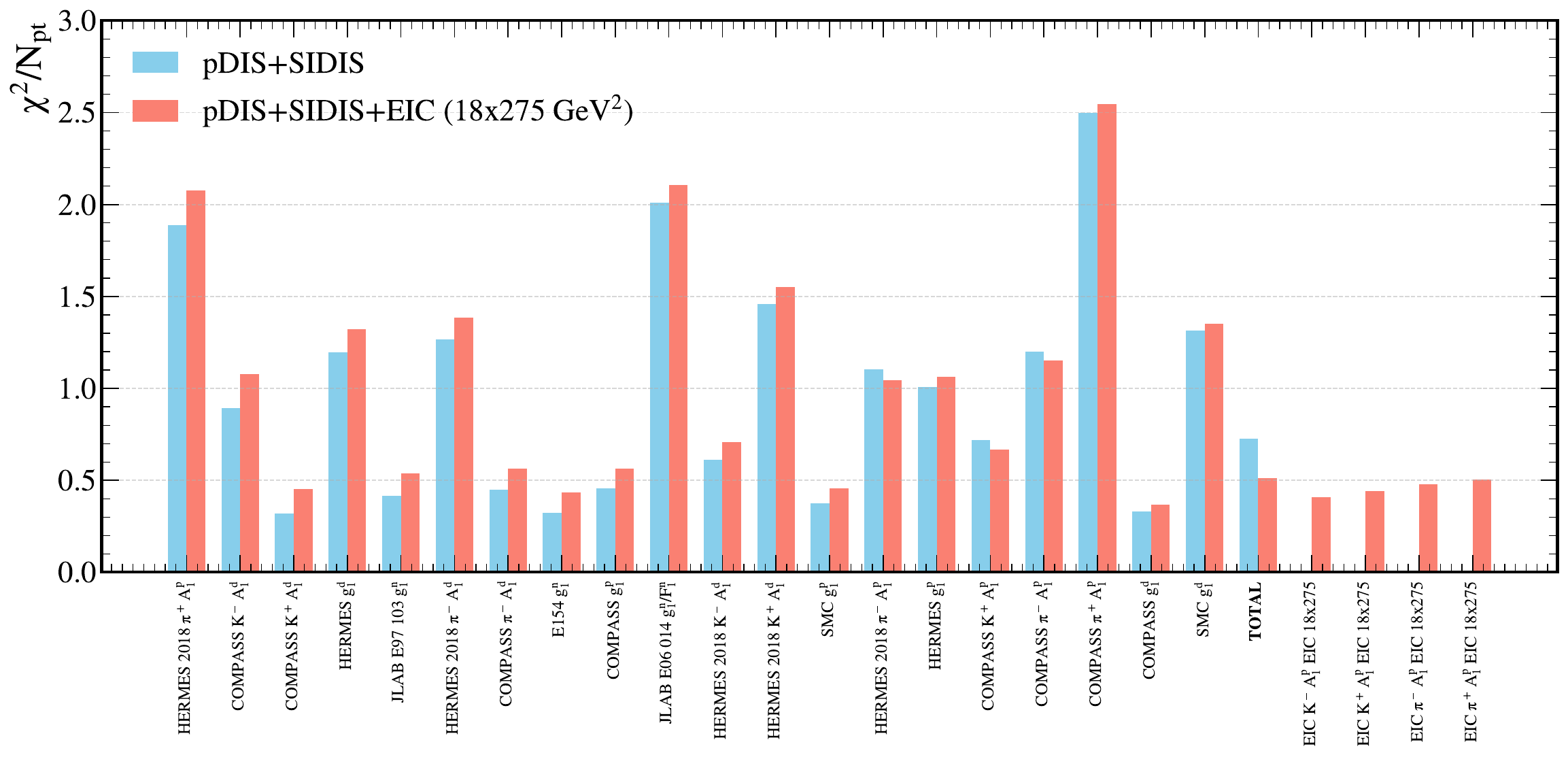}
\begin{center}
\caption{ \small 
Comparison of $\chi^2$ contributions for the fitted data subsets in the {\tt pDIS+SIDIS base} fit (blue)
and in the EIC-augmented fits (red), shown separately for the $5 \times 41~\mathrm{GeV^2}$ (upper) and $18 \times 275~\mathrm{GeV^2}$ (lower)
EIC configurations. 
Results are displayed for the default $z_{\min}>0.2$ selection.} 
\label{fig:Chi2_comparison_noEIC_EIC}
\end{center}
\end{figure*}

A further consistency check is provided by the dataset-by-dataset comparison in
Fig.~\ref{fig:Chi2_comparison_noEIC_EIC}, where we display the $\chi^2$ contributions of each fitted subset
when the EIC pseudodata are excluded (blue) or included (red). 
Results are displayed for the default $z_{\min}>0.2$ selection. 
The overall pattern is that adding EIC 
pseudodata does not deteriorate the description of the existing world measurements: changes in the
$\chi^2$ contributions are modest and fluctuate around the baseline values, as expected when additional
constraints are introduced primarily in kinematic regions not covered by current data.

%
%
%
%

\subsection{Impact of EIC pseudodata on helicity PDFs}\label{subsec:eic_impact}

Since the projected EIC pseudodata considered in this work correspond exclusively to
semi-inclusive DIS measurements with identified hadrons
$\pi^{+}$, $\pi^{-}$, $K^{+}$, and $K^{-}$ in the final state, 
their primary impact is expected on the flavor-separated sea-quark helicity distributions.
In particular, charge-separated pion asymmetries provide strong constraints on the light
antiquark distributions $\Delta\bar u(x)$ and $\Delta\bar d(x)$, while kaon production
offers a unique sensitivity to the polarized strange-quark sector, $\Delta s(x)$ and
$\Delta\bar s(x)$.
The valence polarized PDFs are already relatively well constrained by
existing inclusive DIS data, and the gluon helicity distribution is affected only indirectly
through NLO QCD evolution and PDF correlations.  
We first emphasize a sea-quark--focused presentation, showing 
$\Delta \bar u(x)$, $\Delta d(x)$, and $\Delta s(x)$ (or alternatively $\Delta\bar s(x)$),
which provides the cleanest demonstration of the enhanced sensitivity to the polarized
sea afforded by pion and kaon SIDIS measurements at the EIC. 
As a complementary perspective, one also highlight the role of EIC data in enabling
full flavor separation by presenting $\Delta \bar u(x)$ and $\Delta \bar d(x)$ together with
the corresponding valence distributions $\Delta u(x)$ and $\Delta d(x)$. 
Finally, given the unique role of kaon production, a dedicated focus on the strange sector
can be achieved by showing $\Delta s(x)$, $\Delta\bar s(x)$, and the 
truncated quark-singlet helicity $\Delta \Sigma(Q^2)$, directly addressing the long-standing question of the 
polarized strange-quark contribution to the proton spin.

\subsubsection{EIC at \texorpdfstring{$5 \times 41$}{5x41}~GeV$^2$}\label{subsubsec:eic_5x41} 

\begin{figure*}[htb]
\vspace{0.50cm}
\centering
\includegraphics[width=0.32\textwidth]{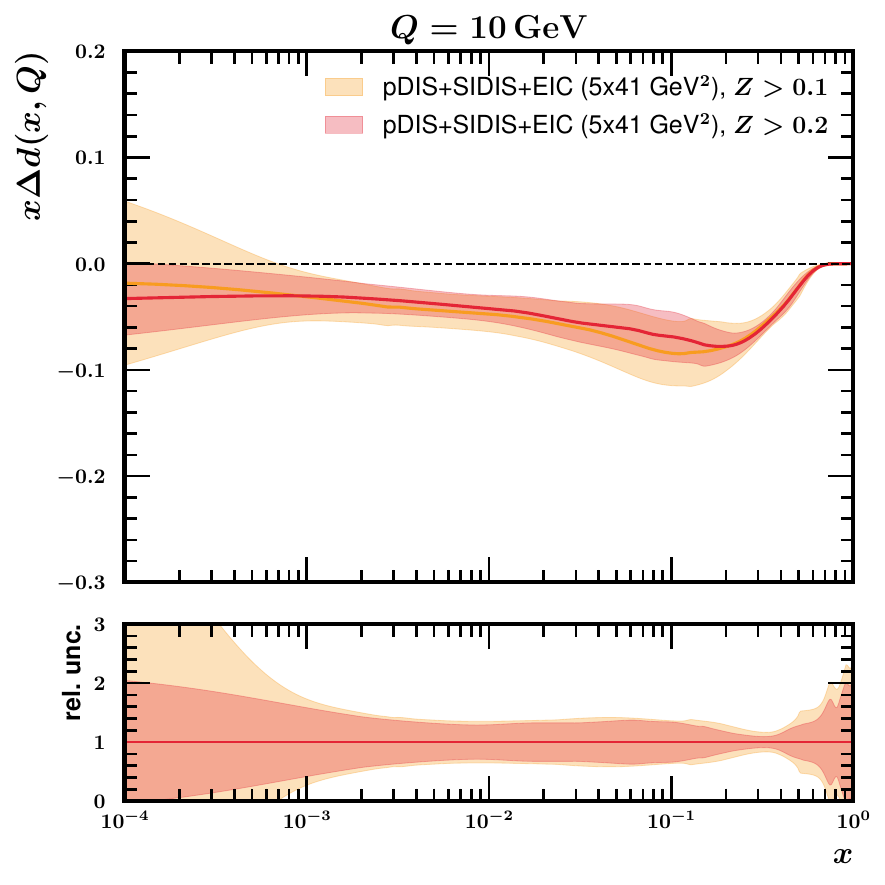} 
\includegraphics[width=0.32\textwidth]{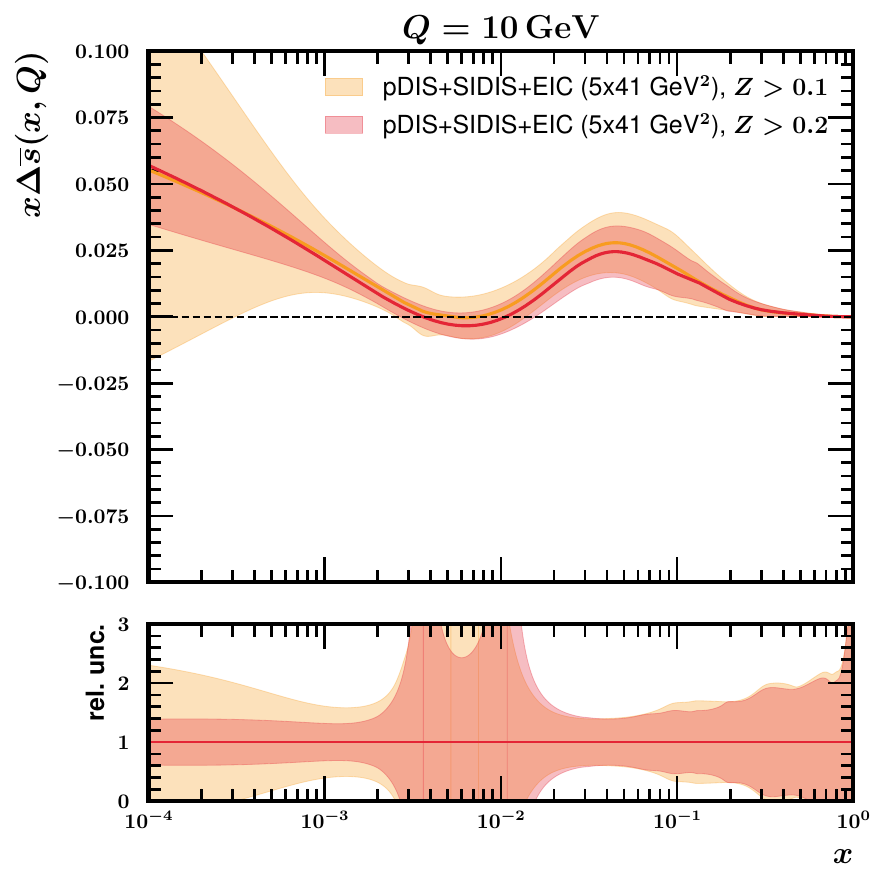}   \\
\includegraphics[width=0.32\textwidth]{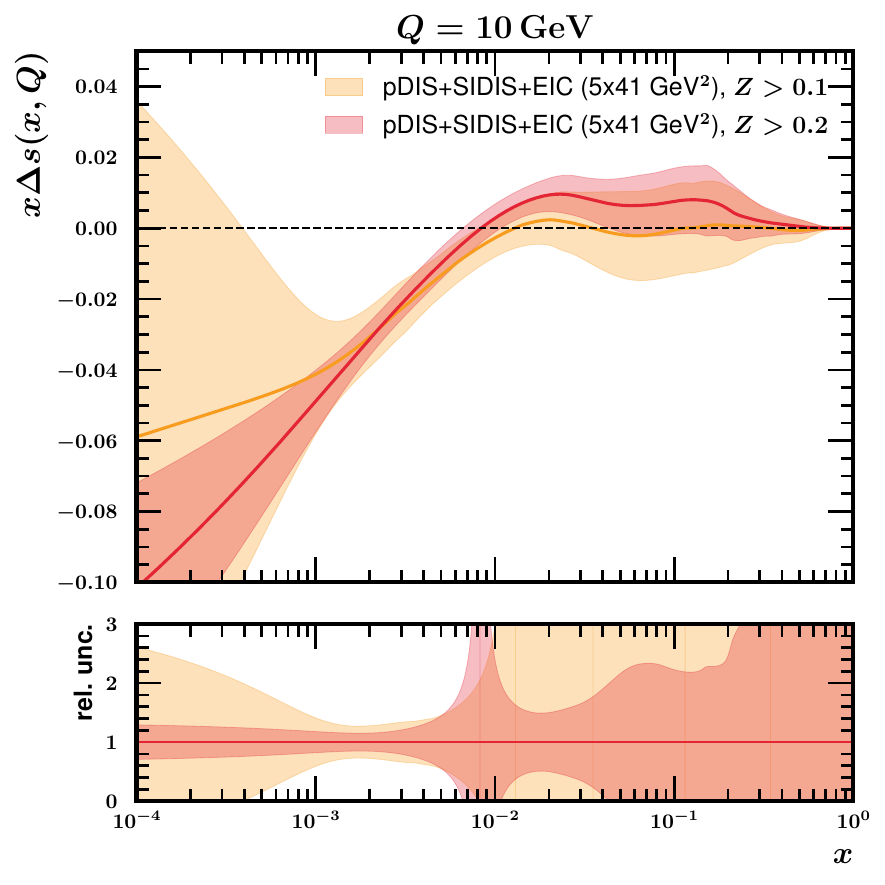}     
\includegraphics[width=0.32\textwidth]{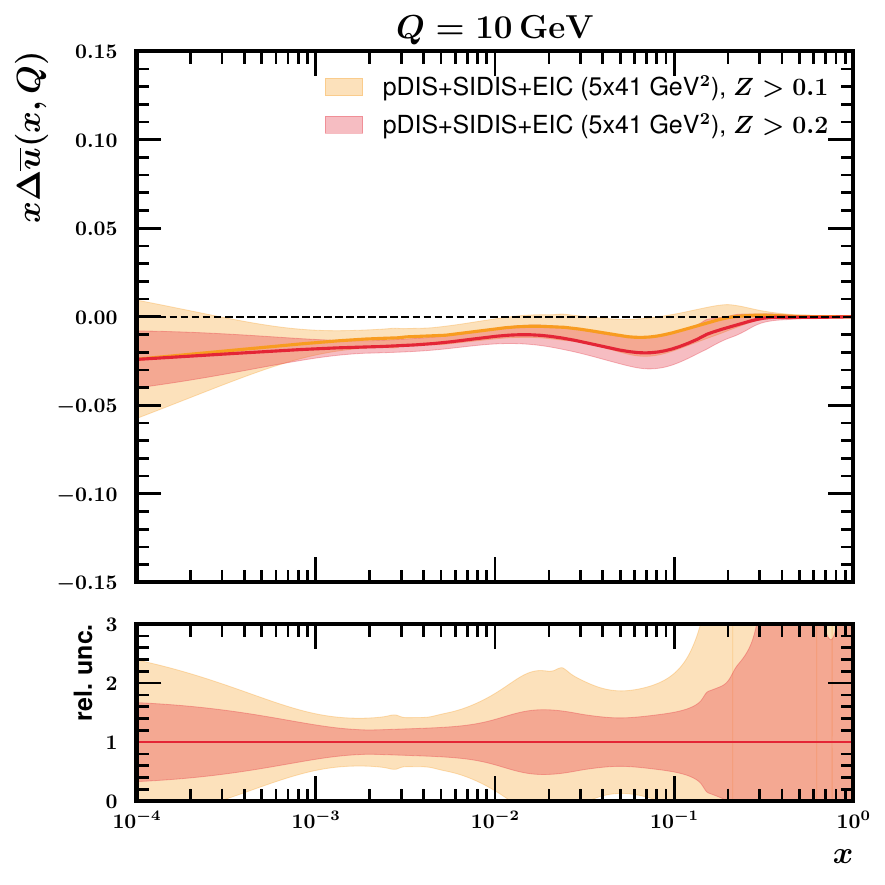}   \\
\begin{center}
\caption{ \small 
Comparison of polarized PDFs at $Q = 10~\mathrm{GeV}$ obtained with EIC pseudodata at $5\times41~\mathrm{GeV^2}$
using two minimum-$z$ selections, $z_{\min}>0.1$ and $z_{\min}>0.2$.}
\label{fig:compare_z_28GeV_Z01_Z02}
\end{center}
\end{figure*}

We first assess the impact of adding the $5 \times41 ~\mathrm{GeV}^2$ EIC pseudodata. 
Figure~\ref{fig:compare_z_28GeV_Z01_Z02} compares representative polarized PDFs at $Q=10~\mathrm{GeV}$
obtained with two alternative selections, $z_{\min}>0.1$ and $z_{\min}>0.2$. 
We adopt, however, a minimal and representative set of
distributions that best illustrate the physics impact of the EIC SIDIS pseudodata, namely the  
sea-quark--focused presentation, by showing $\Delta \bar u(x)$, $\Delta d(x)$, and $\Delta s(x)$ (or alternatively $\Delta\bar s(x)$), 
While the central values are broadly consistent between the two choices, the more conservative cut
$z_{\min}>0.2$ typically leads to a tighter and smoother determination in the kinematic region populated
most densely by the EIC pseudodata, and it is therefore adopted as the default.   
The fact that the $z_{\min}>0.2$ selection can yield smaller PDF uncertainties than the looser cut $z_{\min}>0.1$ reflects the fact that, 
in a global SIDIS fit, more pseudodata points do not automatically imply a stronger effective constraint.  
The lower-$z$ region is theoretically less clean and more sensitive to residual hadronization effects and to the present 
level of fragmentation function uncertainties, since SIDIS asymmetries depend explicitly on both helicity PDFs and FFs. 
Consequently, the additional bins included for $z_{\min}>0.1$ may carry weaker or less stable constraining power, which can broaden 
the spread of fitted replicas.  

As can be seen from Fig.~\ref{fig:compare_z_28GeV_Z01_Z02}, the most visible impact is observed for 
sea-quark distributions that are primarily constrained by SIDIS,
and for the gluon distribution at small $x$, reflecting the combination of SIDIS flavor tagging and the
extended evolution lever arm provided by EIC kinematics. 

%
%
%
%

From a physics standpoint, these trends follow from two EIC features. 
The extended $(x,Q^2)$ lever arm improves sensitivity to $\Delta g(x,Q^2)$ through scaling violations in
NLO evolution, while charge separation in SIDIS improves the disentangling of sea flavor:
pion channels are especially effective for $\Delta\bar u$ and $\Delta\bar d$, whereas kaon channels enhance
sensitivity to $\Delta s$ (and $\Delta\bar s$), subject to the FF dependence.
A more systematic propagation of FF uncertainties is therefore a natural direction for future work.

\subsubsection{EIC at \texorpdfstring{$18\times275$}{18x275}~GeV$^2$}\label{subsubsec:eic_18x275}

\begin{figure*}[htb]
\vspace{0.50cm}
\centering
\includegraphics[width=0.32\textwidth]{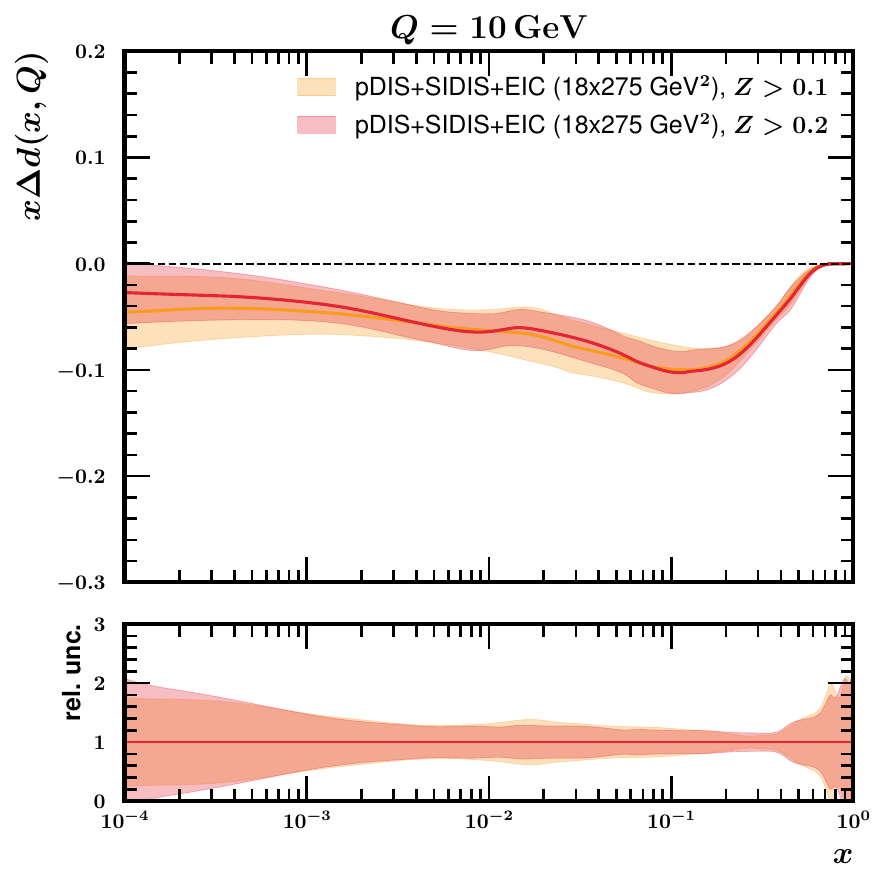}   
\includegraphics[width=0.32\textwidth]{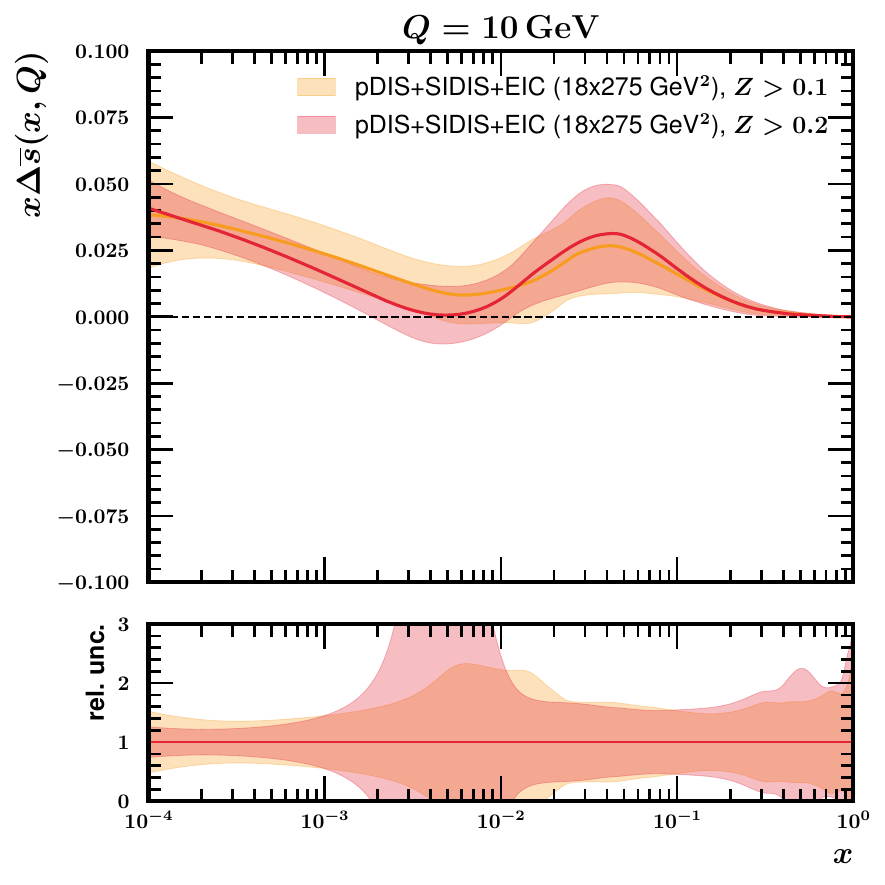} \\  
\includegraphics[width=0.32\textwidth]{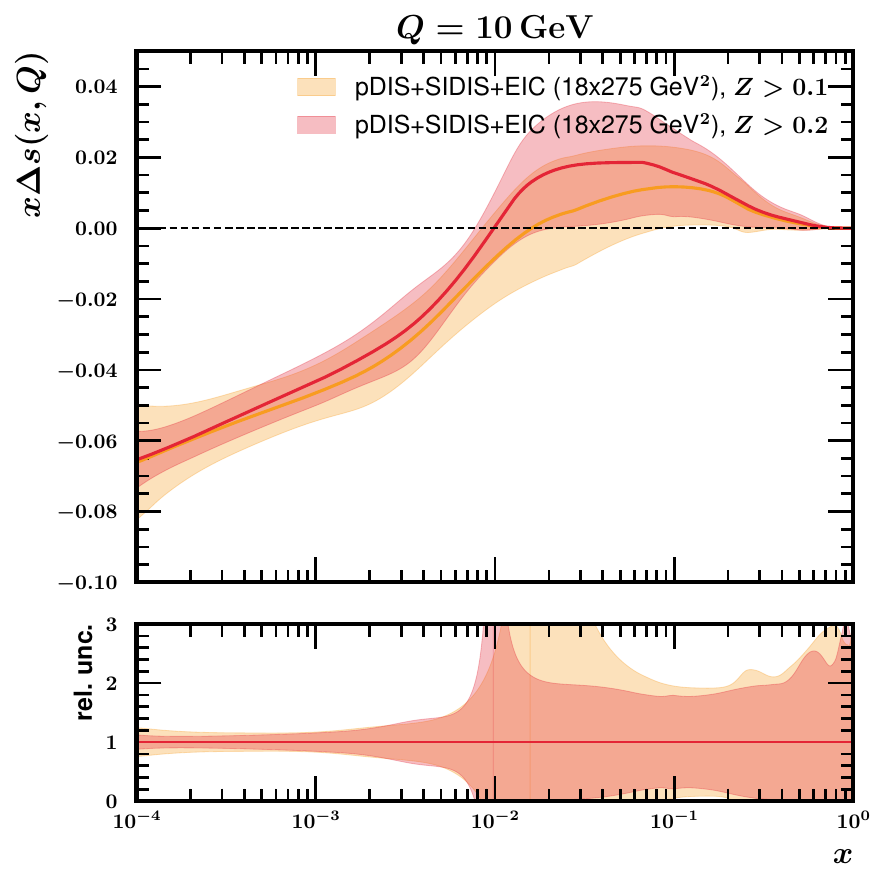}     
\includegraphics[width=0.32\textwidth]{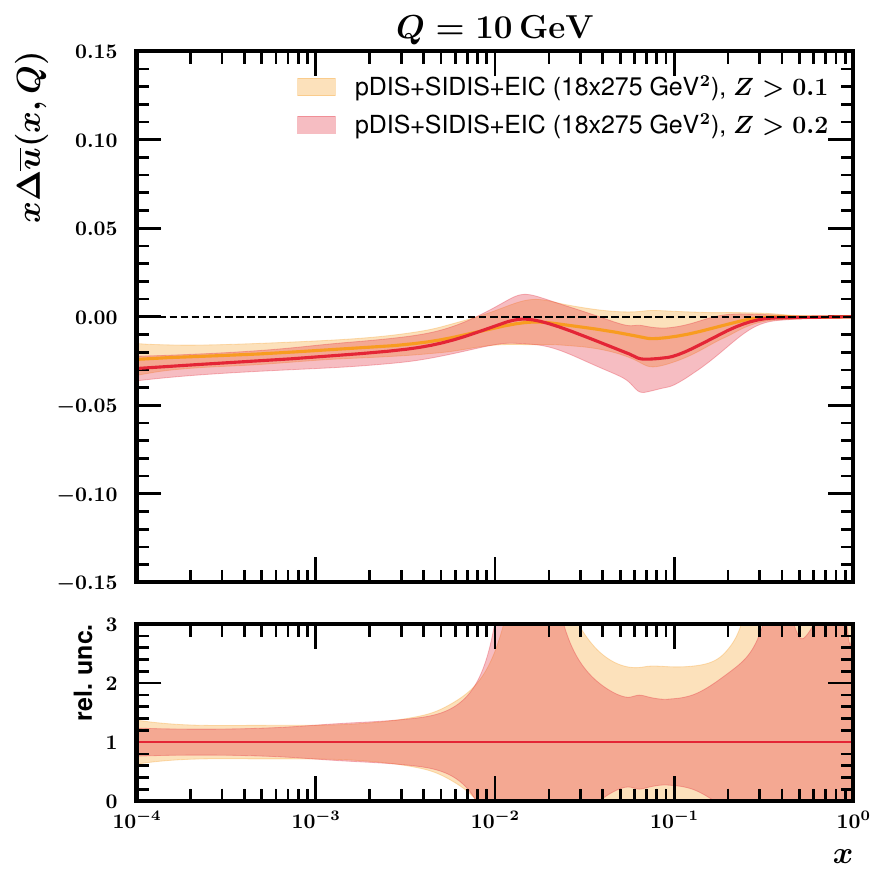}   \\
\begin{center}
\caption{ \small 
Comparison of polarized PDFs at $Q = 10~\mathrm{GeV}$ obtained with EIC pseudodata at $18\times275~\mathrm{GeV^2}$
using two minimum-$z$ selections, $z_{\min}>0.1$ and $z_{\min}>0.2$.}
\label{fig:compare_z_140GeV_Z01_Z02}
\end{center}
\end{figure*}

We next include the higher-energy $18 \times 275 ~\mathrm{GeV^2}$ EIC pseudodata. 
Relative to $5 \times 41~\mathrm{GeV^2}$, this configuration provides a substantially extended reach in $x$ and a
broader $Q^2$ span, strengthening constraints in the small-$x$ region where current data provide limited
direct sensitivity. The dependence on the $z_{\min}$ selection is illustrated in
Fig.~\ref{fig:compare_z_140GeV_Z01_Z02}, again showing that the $z_{\min}>0.2$ choice yields a stable and
typically tighter determination. 

%
%
%
%

\subsection{Comparison with {\tt pDIS+SIDIS base} fit}\label{subsec:comparisons_base_fit}

\begin{figure*}[htb]
\vspace{0.50cm}
\centering
\includegraphics[width=0.32\textwidth]{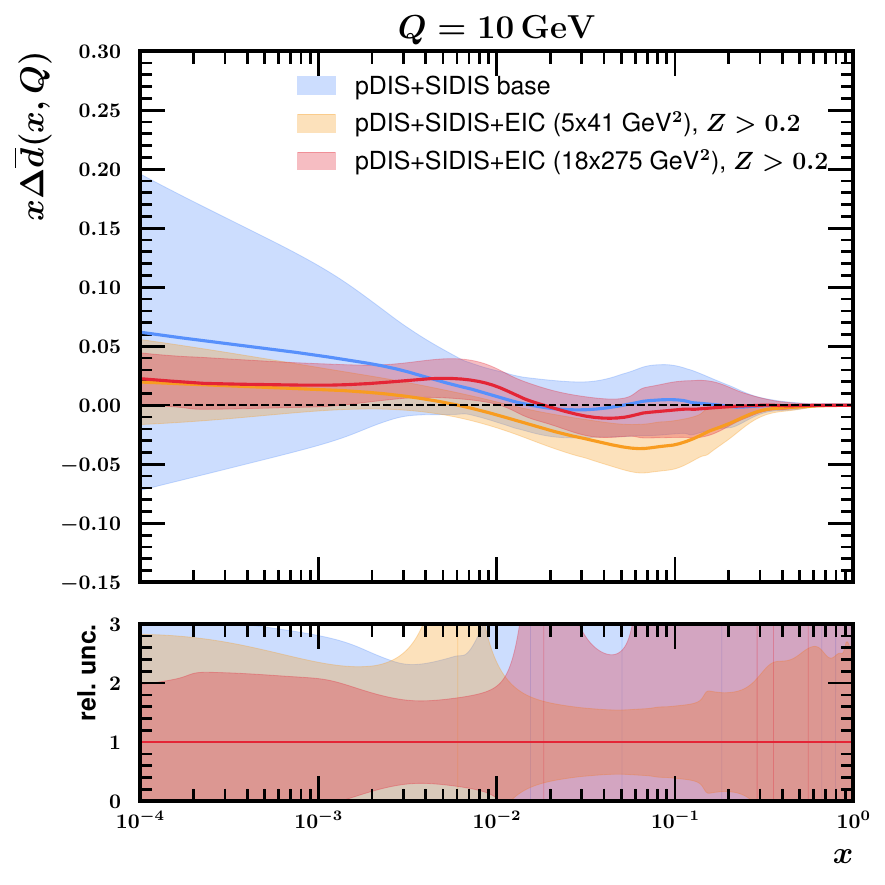} 
\includegraphics[width=0.32\textwidth]{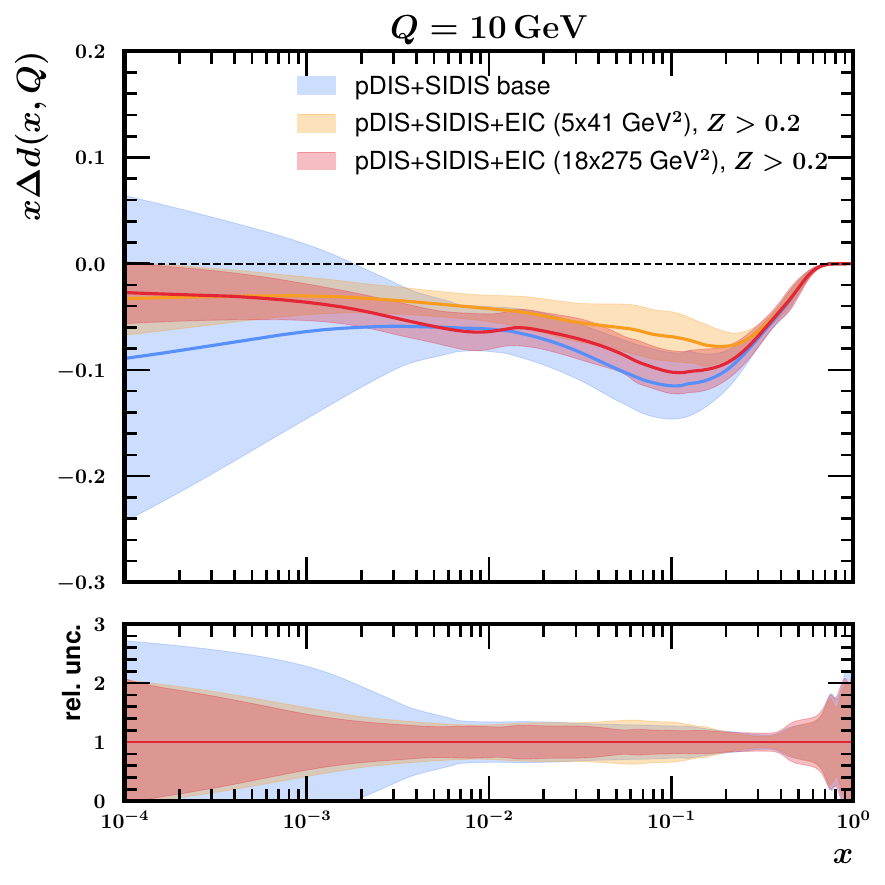}  \\ 
\includegraphics[width=0.32\textwidth]{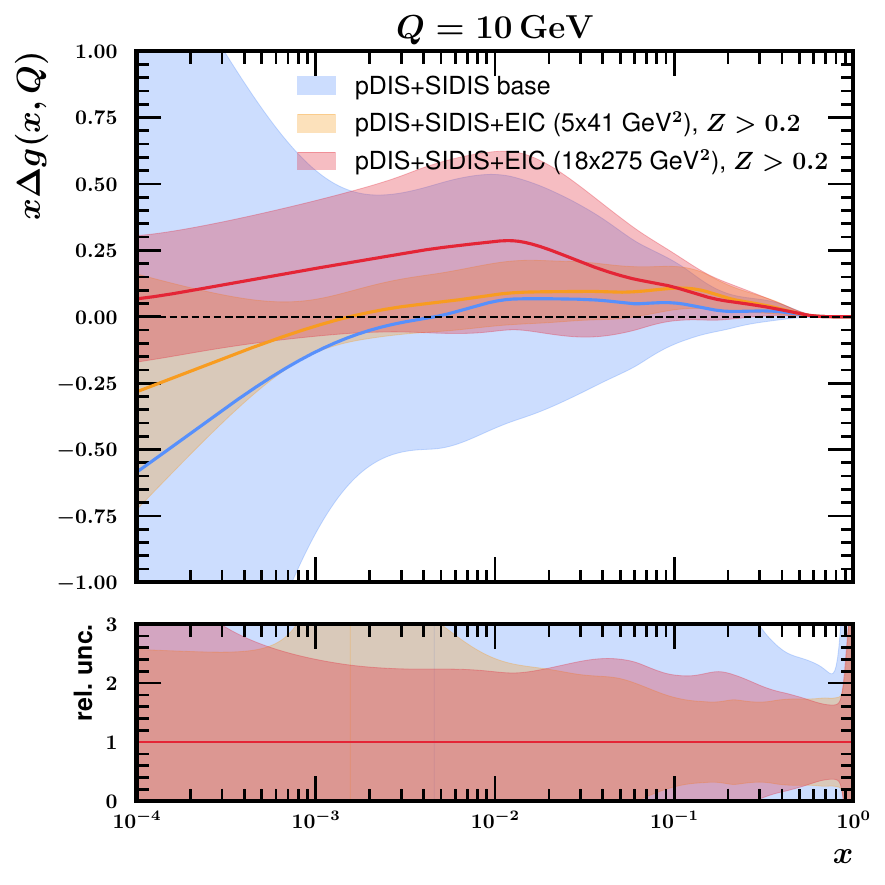}   
\includegraphics[width=0.32\textwidth]{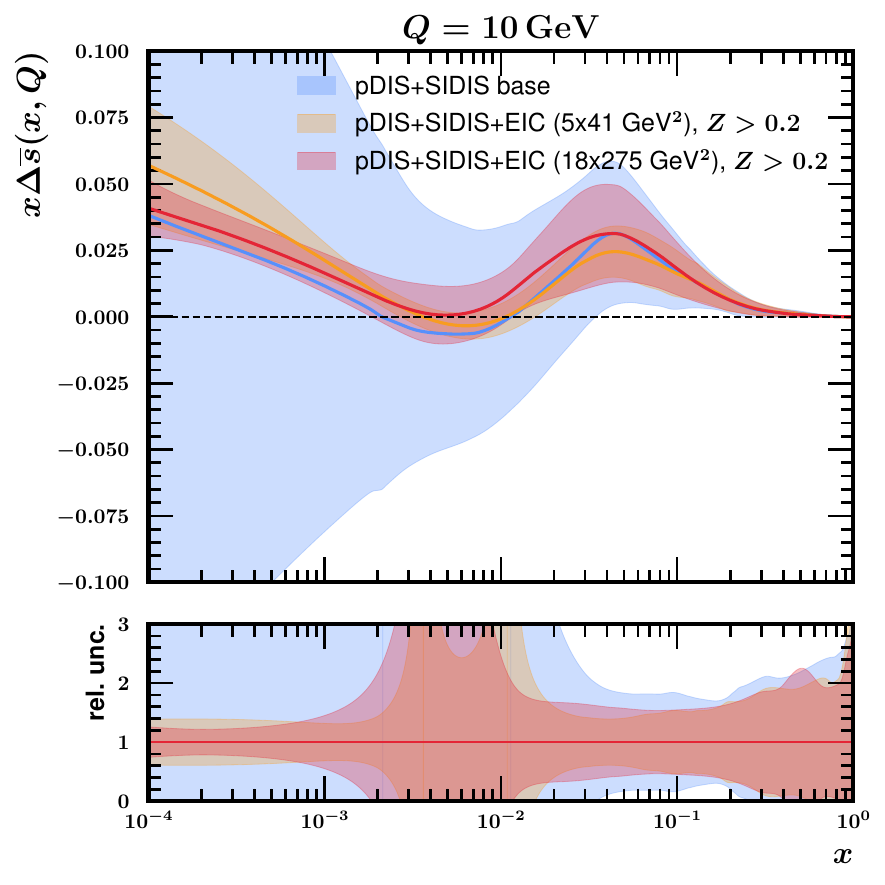}   \\
\includegraphics[width=0.32\textwidth]{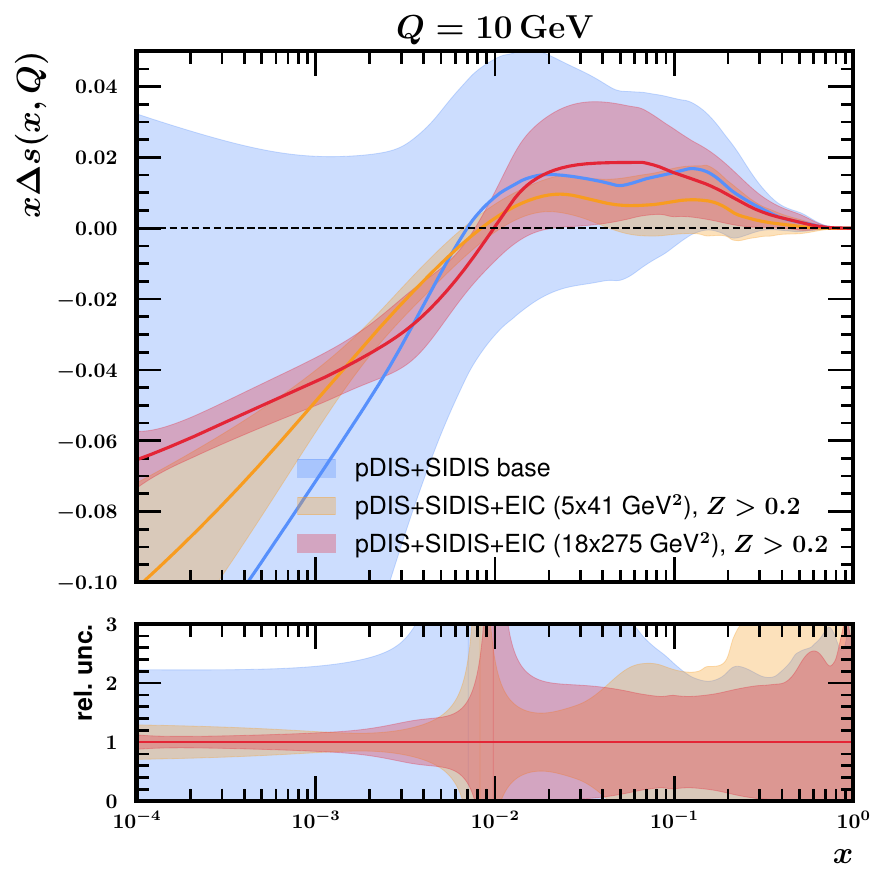}     
\includegraphics[width=0.32\textwidth]{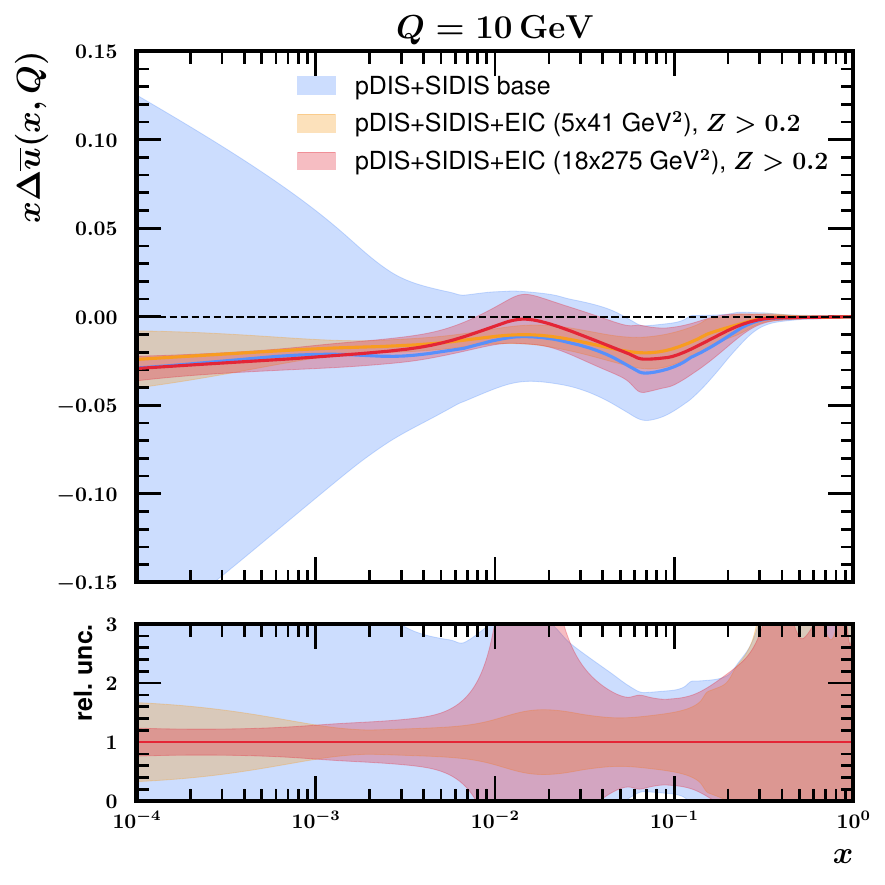}   \\
\includegraphics[width=0.32\textwidth]{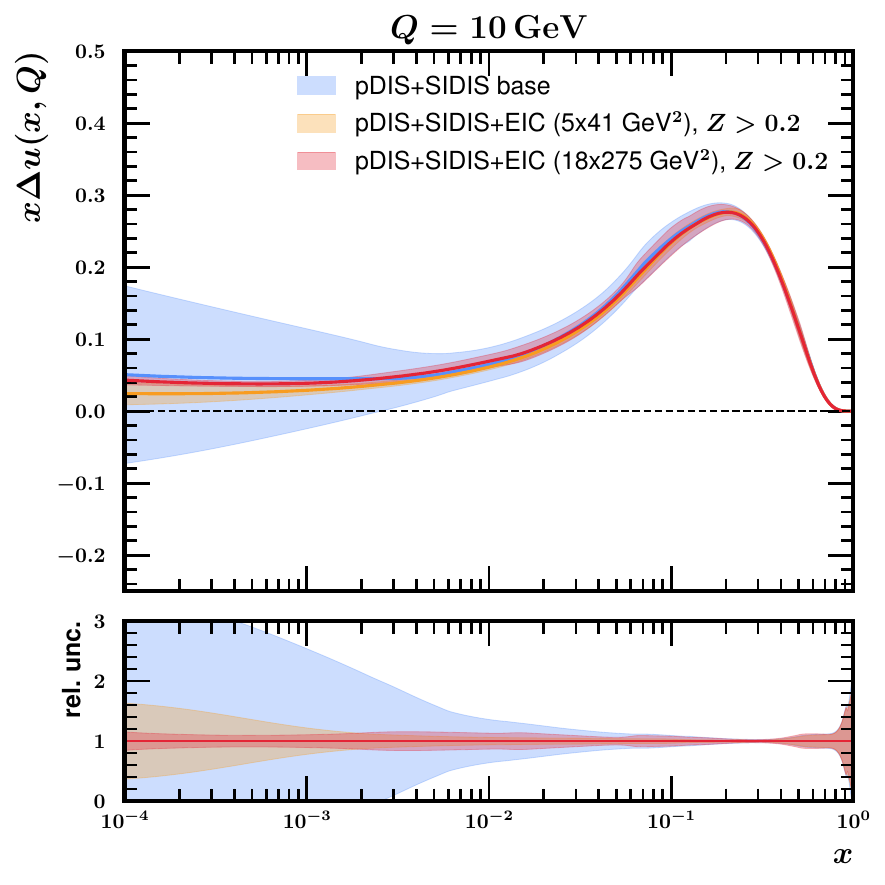}
\begin{center}
\caption{ \small 
Comparison of polarized PDFs at $Q = 10~\mathrm{GeV}$ obtained from the
{\tt pDIS+SIDIS base} fit and from fits including projected EIC SIDIS pseudodata
for the two beam-energy configurations $E_e\times E_p=5\times41~\mathrm{GeV^2}$ and
$18\times275~\mathrm{GeV^2}$, using the default hadron-energy cut $z>z_{\min}=0.2$.}
\label{fig:compare_Z02_Base}
\end{center}
\end{figure*}

To emphasize the net impact of the default EIC setup, Fig.~\ref{fig:compare_Z02_Base} compares the
{\tt pDIS+SIDIS base} fit with the EIC-augmented fit using $z_{\min}>0.2$.
The overall trend is that EIC pseudodata reduce the PDF uncertainties in the regions where EIC data provide
dense coverage in $(x, Q^2)$, thereby reducing extrapolation uncertainties associated with the previously
weakly constrained small-$x$ domain. This effect is particularly evident for SIDIS-driven sea-quark
combinations and for the gluon helicity distribution, which benefits from the extended scaling-violation
information at small $x$. 

%
%
%
%

\subsection{Comparison with recent global determinations}\label{subsec:comparisons}

\begin{figure*}[htb]
\vspace{0.50cm}
\centering
\includegraphics[width=0.32\textwidth]{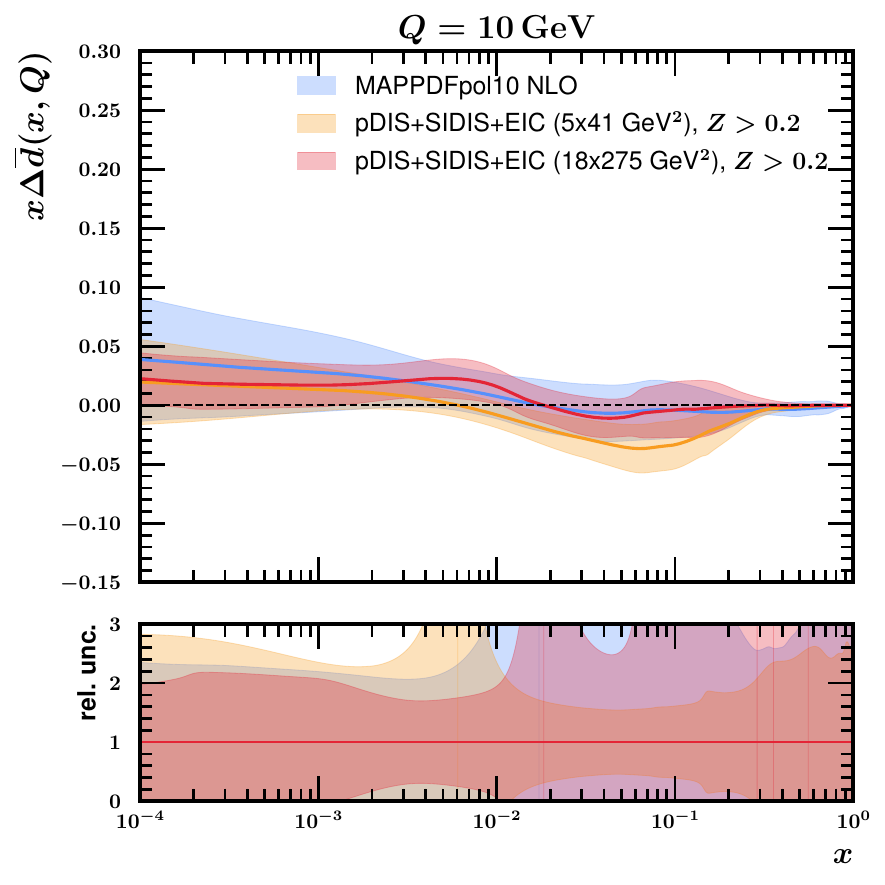} 
\includegraphics[width=0.32\textwidth]{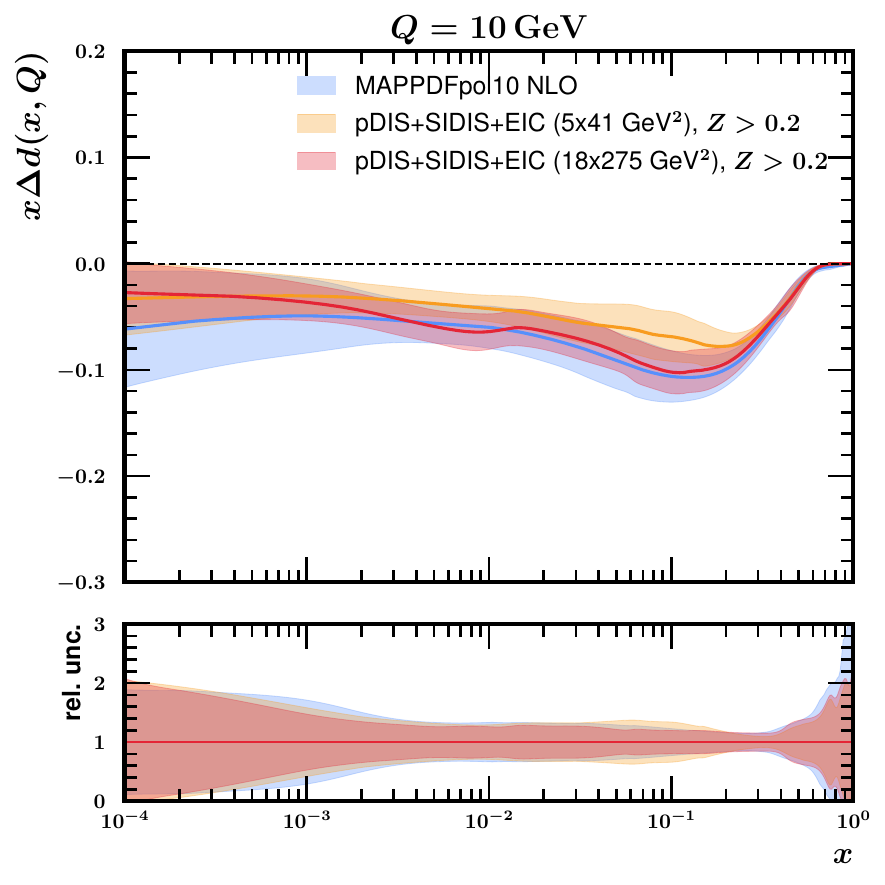}  \\ 
\includegraphics[width=0.32\textwidth]{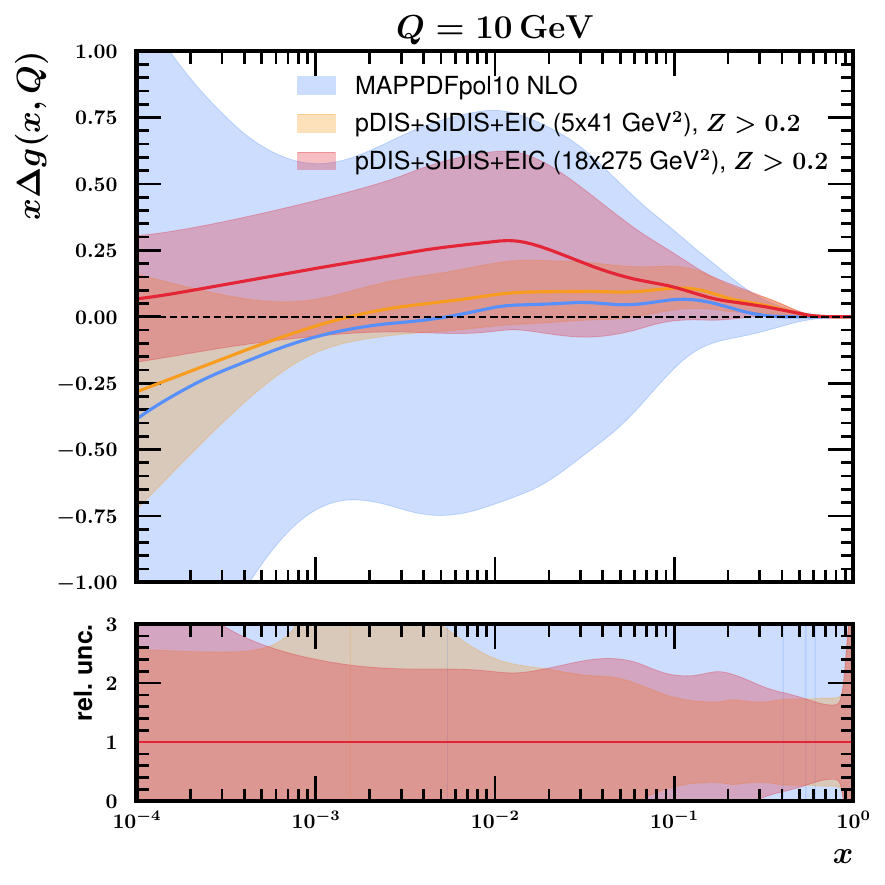}   
\includegraphics[width=0.32\textwidth]{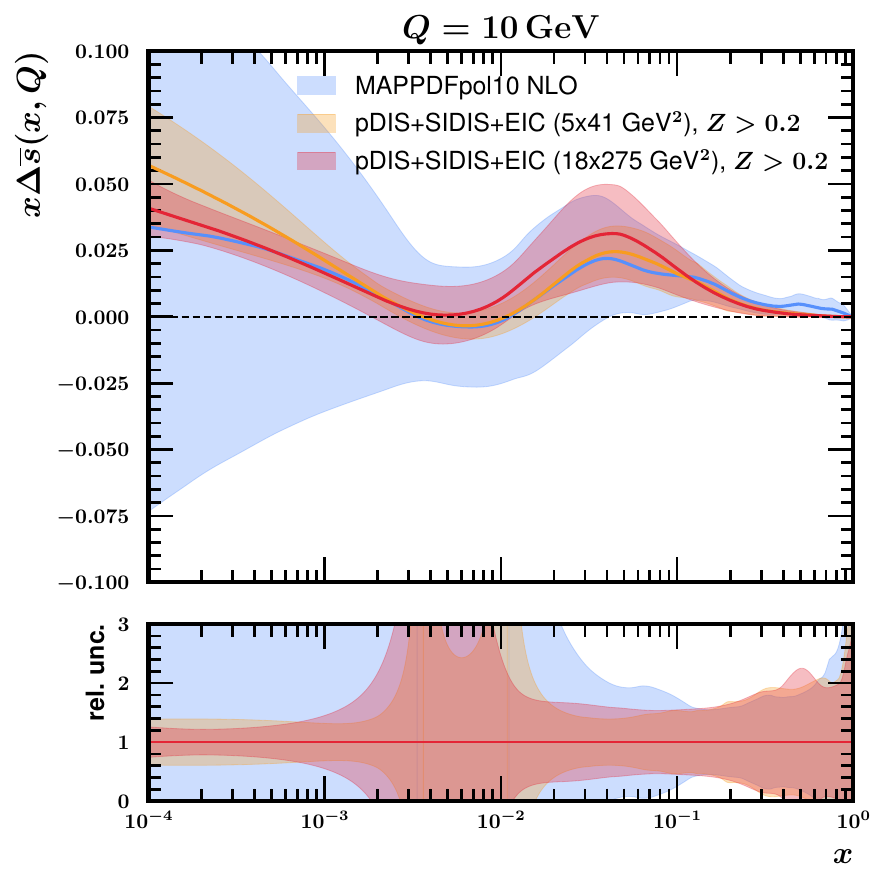}   \\
\includegraphics[width=0.32\textwidth]{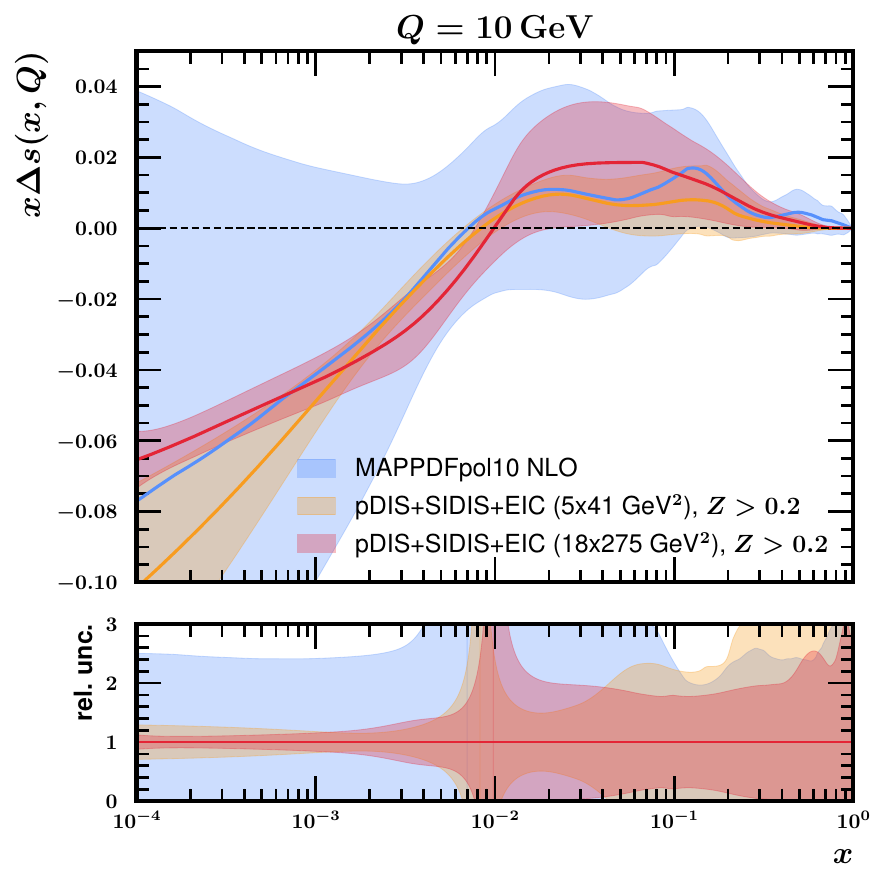}     
\includegraphics[width=0.32\textwidth]{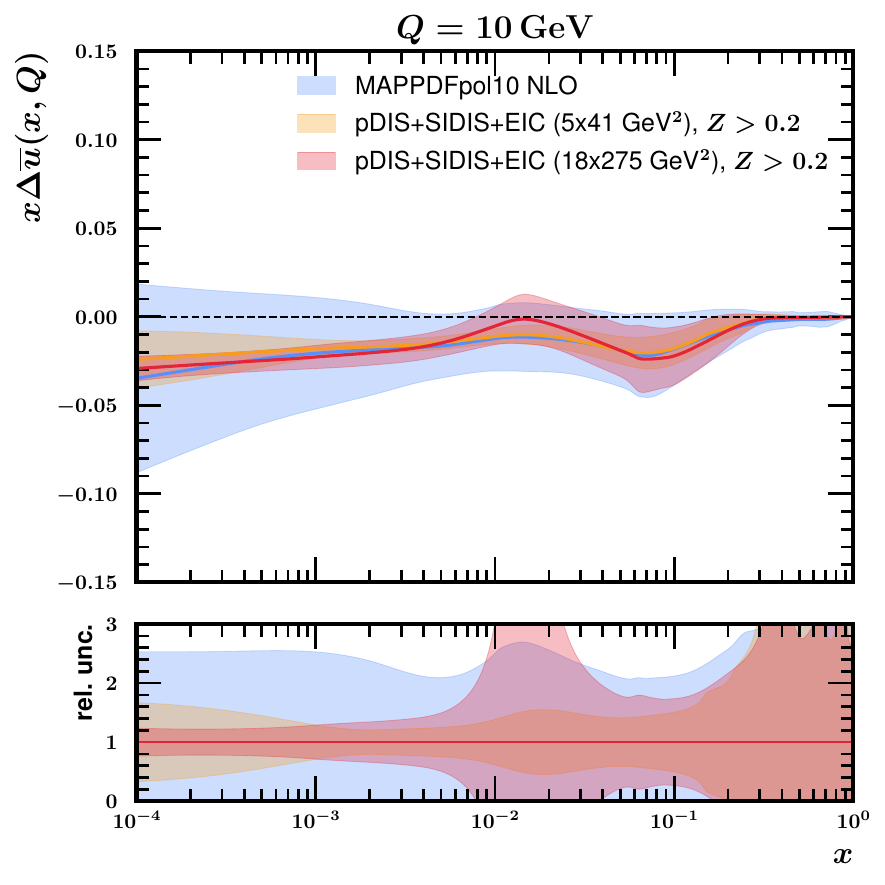}   \\
\includegraphics[width=0.32\textwidth]{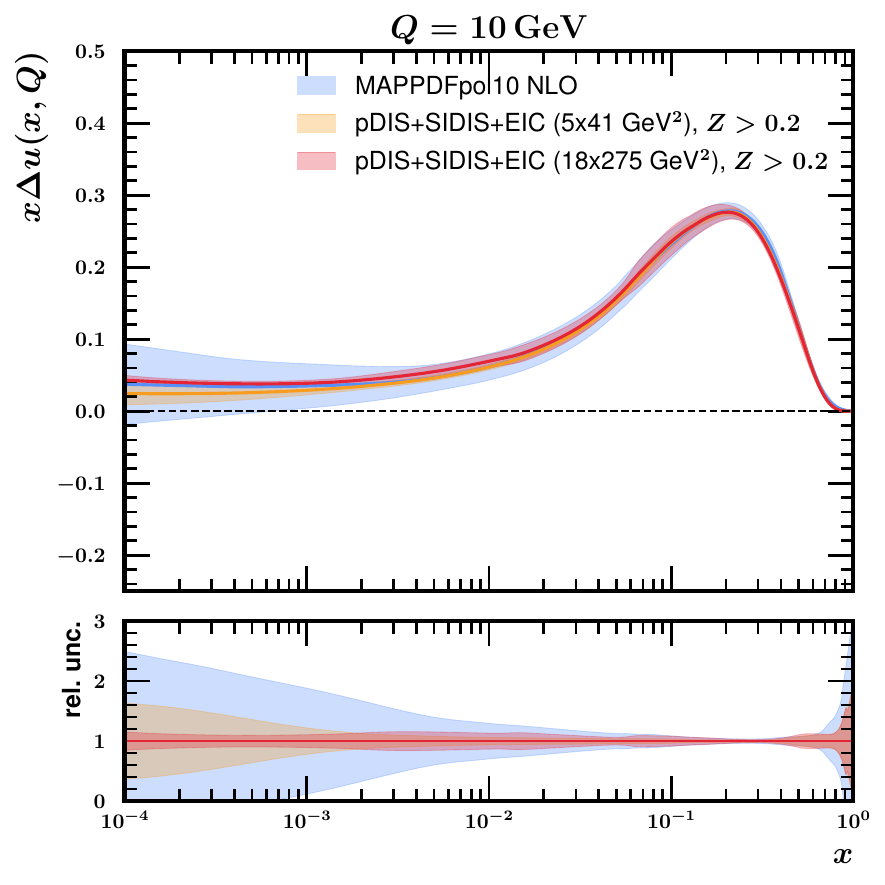}
\begin{center}
\caption{ \small 
Comparison of polarized PDFs at $Q=10~\mathrm{GeV}$ between our default EIC-augmented fit
(with $z_{\min}>0.2$) and the {\tt MAPPDFpol10NLO} determination~\cite{Bertone:2024taw} at NLO accuracy.}
\label{fig:compare_Z02_MAPPDFpol10NLO}
\end{center}
\end{figure*}

Finally, we compare our default EIC-augmented PDFs ($z_{\min}>0.2$) to recent NLO global determinations.
Figure~\ref{fig:compare_Z02_MAPPDFpol10NLO} compares our results to the MAPPDFpol10NLO set~\cite{Bertone:2024taw}.
Overall agreement in the well-constrained region provides a non-trivial validation of the baseline
methodology and dataset treatment. 
Differences are expected given the different analysis choices and, 
in particular, the inclusion in our study of projected EIC pseudodata that enhance constraints at small $x$.

%
%
%
%

\begin{figure*}[htb]
\vspace{0.50cm}
\centering
\includegraphics[width=0.32\textwidth]{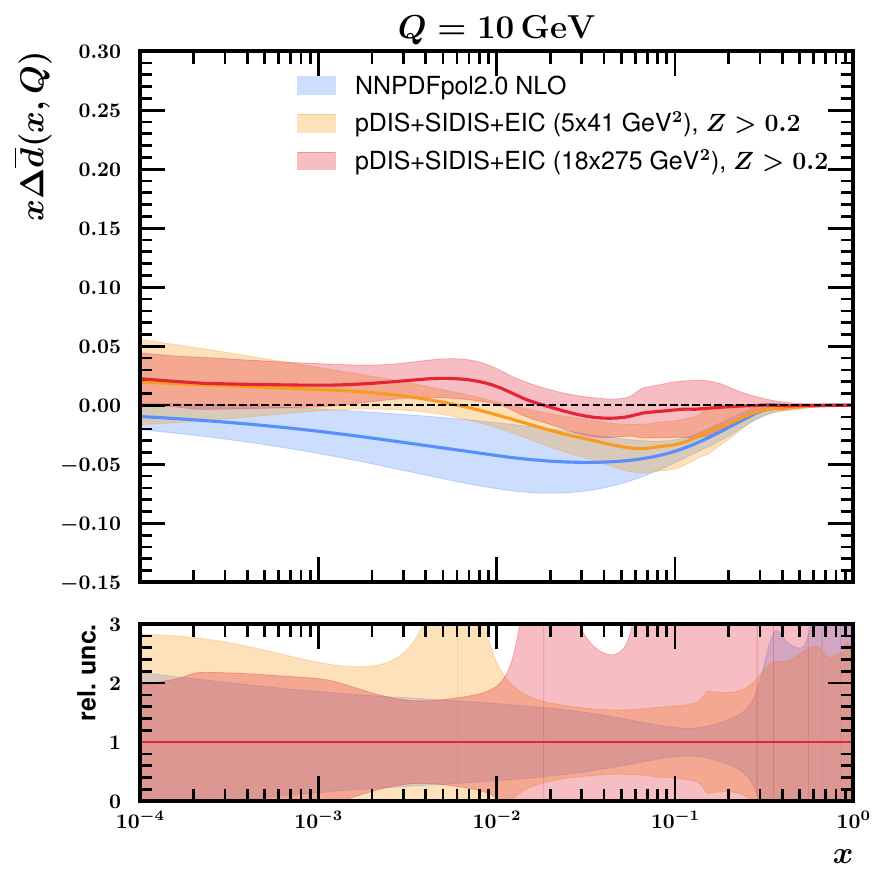} 
\includegraphics[width=0.32\textwidth]{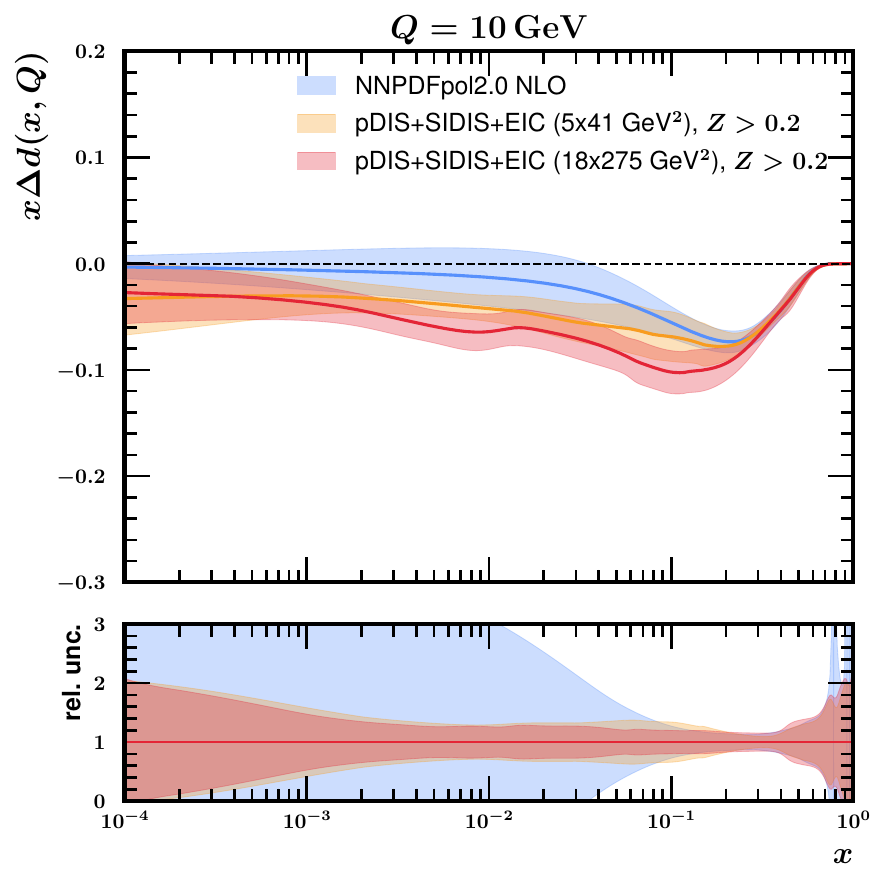}  \\ 
\includegraphics[width=0.32\textwidth]{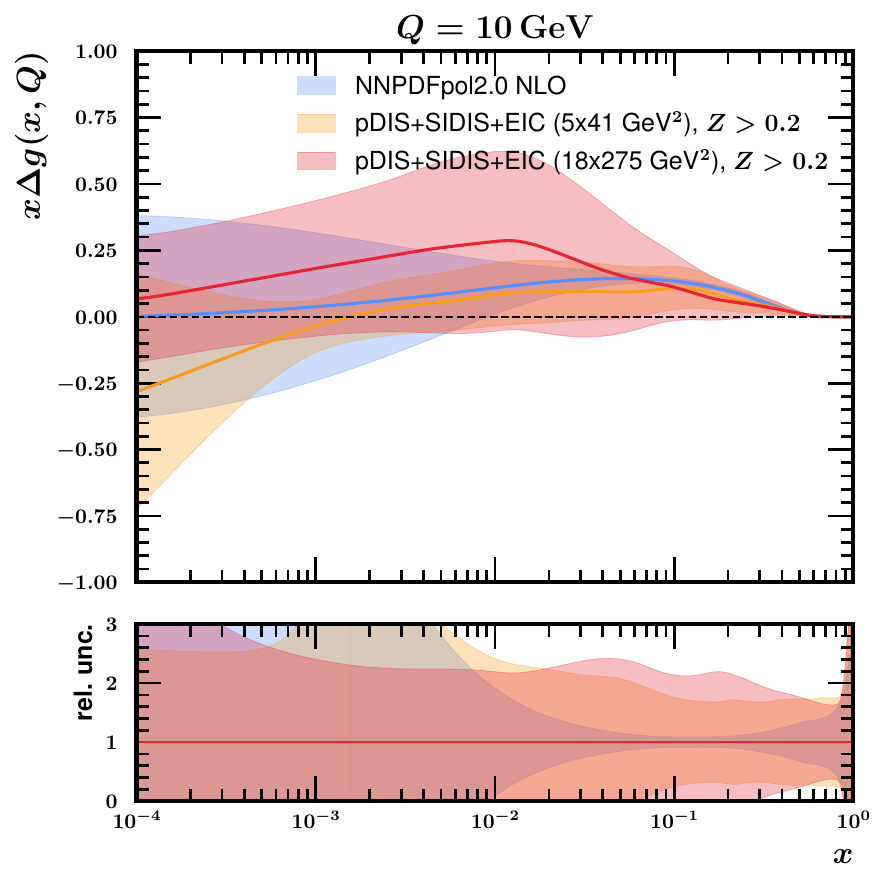}   
\includegraphics[width=0.32\textwidth]{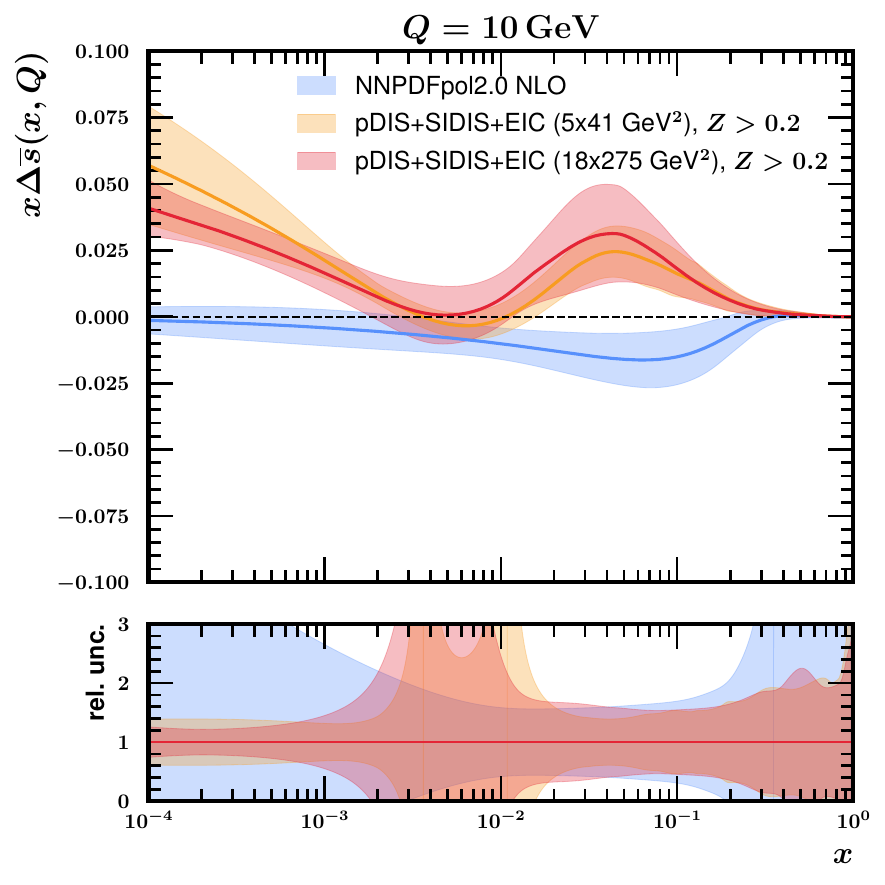}   \\
\includegraphics[width=0.32\textwidth]{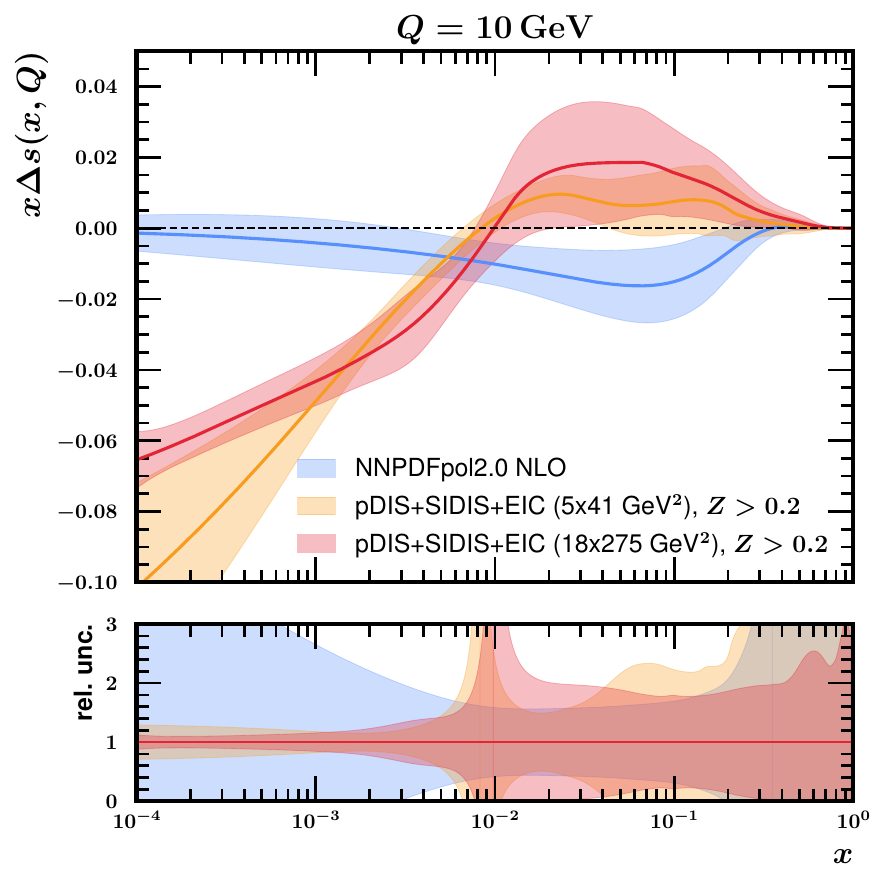}     
\includegraphics[width=0.32\textwidth]{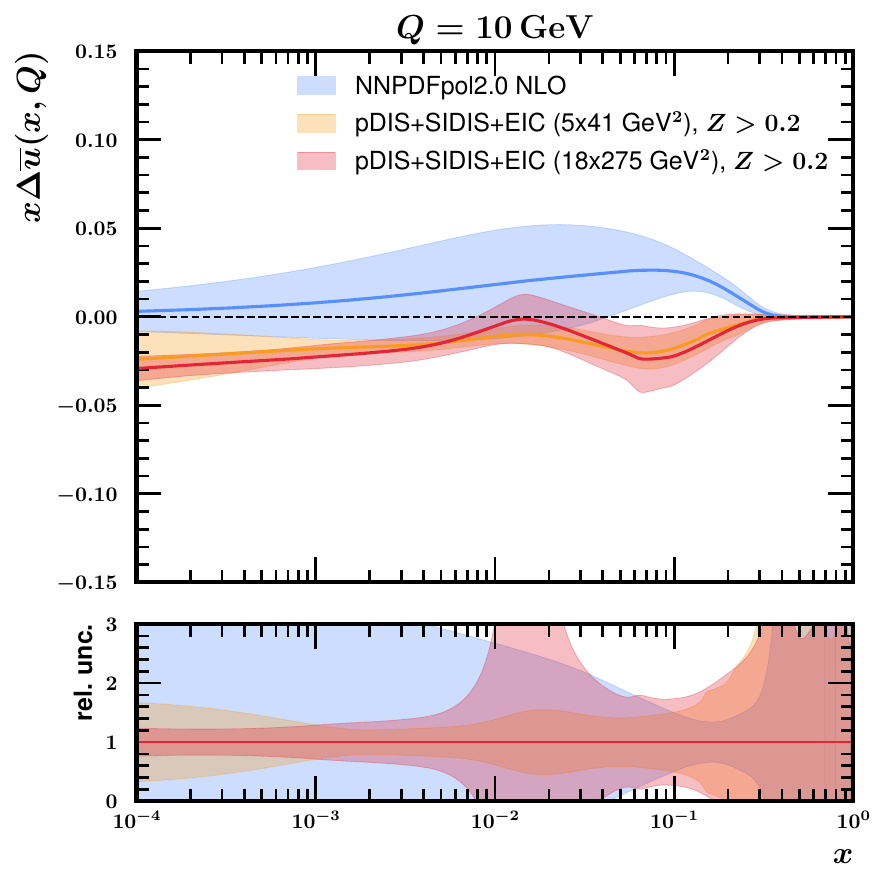}   \\
\includegraphics[width=0.32\textwidth]{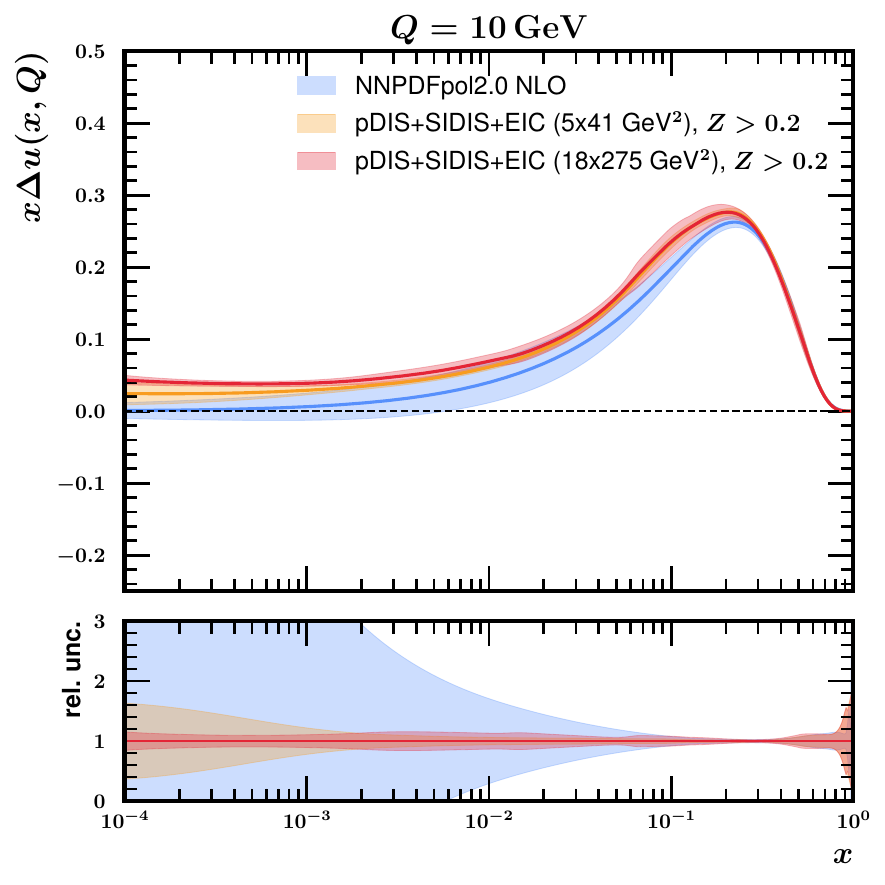}
\begin{center}
\caption{ \small 
Comparison of polarized PDFs at $Q=10~\mathrm{GeV}$ between our default EIC-augmented fit
(with $z_{\min}>0.2$) and the {\tt NNPDFpol2.0} NLO determination~\cite{Cruz-Martinez:2025ahf}.}
\label{fig:Wmass_FMi}
\end{center}
\end{figure*}

A complementary comparison to the NNPDFpol2.0 NLO set~\cite{Cruz-Martinez:2025ahf} is shown in
Fig.~\ref{fig:Wmass_FMi}. While the scope and dataset choices differ between analyses, qualitative
consistency in the data-constrained region is expected. 
In our analysis, the inclusion of charge-identified SIDIS asymmetries for $\pi^+$, $\pi^-$, $K^+$  and $K^-$ from
HERMES and COMPASS, supplemented by EIC SIDIS projections, provides direct sensitivity to flavor
separation in the light sea and, crucially, to strangeness through the kaon channels. This additional
information is reflected in the significantly reduced uncertainty bands at medium-to-small $x$ and in
shape differences that are most visible for the strange sector. In particular, while {\tt NNPDFpol2.0} assumes
$\Delta s(x,Q_0)=\Delta\bar s(x,Q_0)$ at the parameterization scale, we allow $\Delta s$ and $\Delta\bar s$ to be fitted
independently; with charge-separated kaon observables, the EIC projections further enhance sensitivity
to a possible $\Delta s$-$\Delta \bar s$ helicity asymmetry and lead to a noticeably improved constraint on both
$x\,\Delta s$ and $x\,\Delta\bar s$ in the region where current data provide only limited leverage.
In regions where present measurements provide limited constraints, the inclusion of EIC pseudodata in
our study therefore plays a central role in reducing uncertainties, highlighting the potential impact of
future EIC SIDIS measurements on the determination of polarized PDFs, and in particular on the
strange helicity density and its possible quark-antiquark asymmetry.  

%
%
%
%

Overall, these results indicate that the inclusion of EIC SIDIS pseudodata preserves a stable description of the existing polarized DIS and SIDIS measurements, as seen from the largely unchanged fit quality across datasets. The main phenomenological outcome is the significant increase in constraining power, especially for SIDIS-sensitive sea-quark helicity PDFs and for the gluon helicity distribution at small $x$. In the present setup, the more conservative selection $z_{\min}>0.2$ also leads to the most stable and, in several cases, the tightest PDF determination.

To quantify the impact of EIC pseudo-data on the quark and gluon spin contributions,
we study the truncated first moments of the quark-singlet and gluon helicity distributions,

\begin{align}
M_{\Sigma}(x_{\min}) &= \int_{x_{\min}}^{1}\!dx\,\Delta\Sigma(x,Q^{2})\,,\\
M_{g}(x_{\min}) &= \int_{x_{\min}}^{1}\!dx\,\Delta g(x,Q^{2})\,,\\
M_{\mathrm{spin}}(x_{\min}) &= \int_{x_{\min}}^{1}\!dx\,\left[\tfrac12\,\Delta\Sigma(x,Q^{2})+\Delta g(x,Q^{2})\right]\,,
\end{align}

evaluated at $Q^{2}=100~\mathrm{GeV}^{2}$.  
Here $\Delta \Sigma = \sum_{q}\left(\Delta q+ \Delta\bar q\right)$ denotes the quark-singlet
helicity combination, with the sum taken over the active quark flavor in the ZM-VFNS
(at this scale including $u,d,s$ and charm, and, where applicable, bottom).

\begin{figure*}[htb]
\vspace{0.50cm}
\centering
\includegraphics[width=0.6\textwidth]{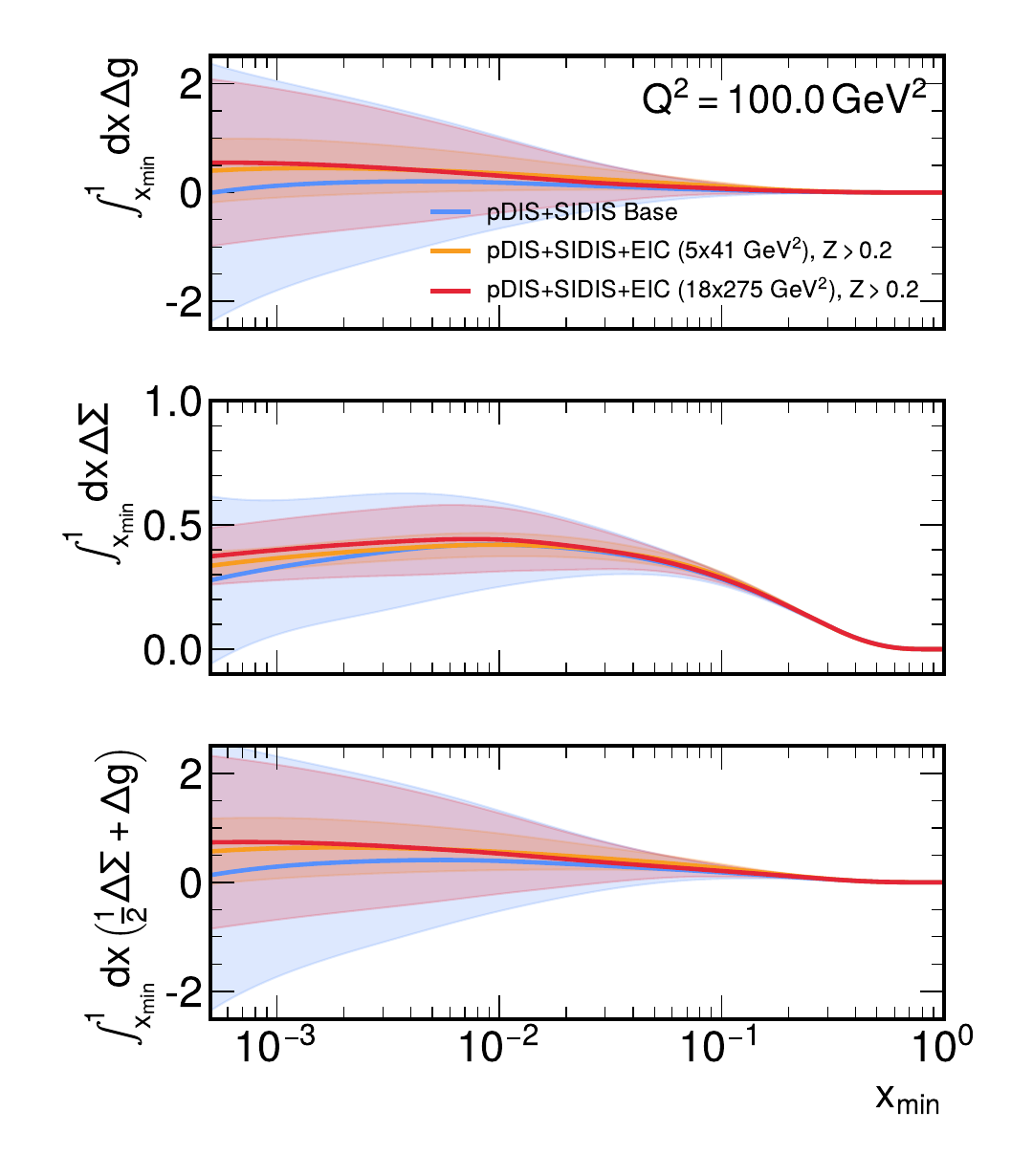} 
\begin{center}
\caption{\small
Truncated first moments of the gluon and quark-singlet polarized PDFs at $Q^{2}=100~\mathrm{GeV}^{2}$ as 
functions of the lower integration limit $x_{\min}$.
Shown are the truncated integrals
$M_{g}(x_{\min})=\int_{x_{\min}}^{1}\!dx\,\Delta g(x,Q^{2})$ (top),
$M_{\Sigma}(x_{\min})=\int_{x_{\min}}^{1}\!dx\,\Delta\Sigma(x,Q^{2})$ (middle), and
$M_{\mathrm{spin}}(x_{\min})=\int_{x_{\min}}^{1}\!dx\,\big[\tfrac12\,\Delta\Sigma(x,Q^{2})+\Delta g(x,Q^{2})\big]$ (bottom),
where $\Delta\Sigma=\sum_{q}\left(\Delta q+\Delta\bar q\right)$ is summed over the active 
quark flavors in the ZM-VFNS.
The solid curves correspond to the central member (replica $0$) of each LHAPDF set, while 
the shaded bands represent the $1\sigma$ uncertainties obtained from the Monte Carlo replica ensemble.
We compare the baseline fit ({\tt pDIS+SIDIS Base}) with two fits including EIC pseudo-data at $(5\times 41)$~GeV$^2$ 
and $(18\times 275)$~GeV$^2$ (both with $z>0.2$).}
\label{fig:truncated_moments}
\end{center}
\end{figure*}

Figure~\ref{fig:truncated_moments} shows these moments as functions of the lower integration
limit $x_{\min}$ for our baseline {\tt pDIS+SIDIS} QCD fit and for fits augmented by EIC pseudo-data
at $5\times 41$~GeV$^2$ and $18\times 275$~GeV$^2$ (with $z>0.2$).
The baseline fit exhibits rapidly growing uncertainties as $x_{\min}$ is lowered,
reflecting the limited present-day constraints on polarized PDFs in the small-$x$ region,
most notably for $\Delta g$.
Upon including EIC information, the uncertainty bands shrink substantially across all panels,
with the strongest improvement at small $x_{\min}$ where extrapolation effects are otherwise dominant.
In particular, the reduction of the uncertainty on $M_{g}(x_{\min})$ propagates directly to the
combination $M_{\mathrm{spin}}(x_{\min})$, highlighting that EIC measurements provide decisive
leverage on the low-$x$ contribution to the gluon helicity and thus sharpen the determination
of the quark-plus-gluon helicity contribution entering the proton spin sum rule.
At larger $x_{\min}$, the moments become increasingly stable and the three fits converge, consistent
with existing fixed-target constraints at moderate and large $x$.

%
%
%
%

\begin{figure*}[htb]
\vspace{0.50cm}
\centering
\includegraphics[width=0.48\textwidth]{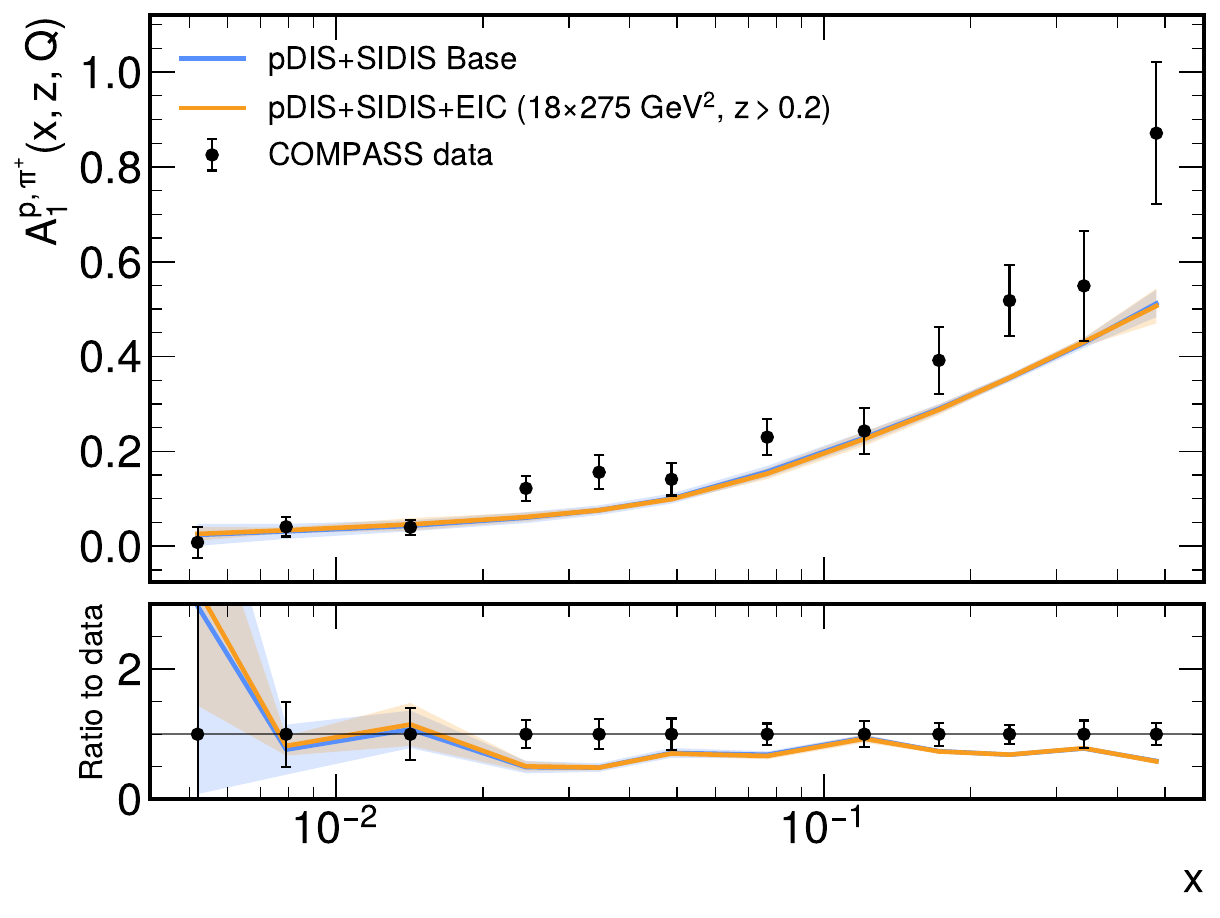} 
\includegraphics[width=0.48\textwidth]{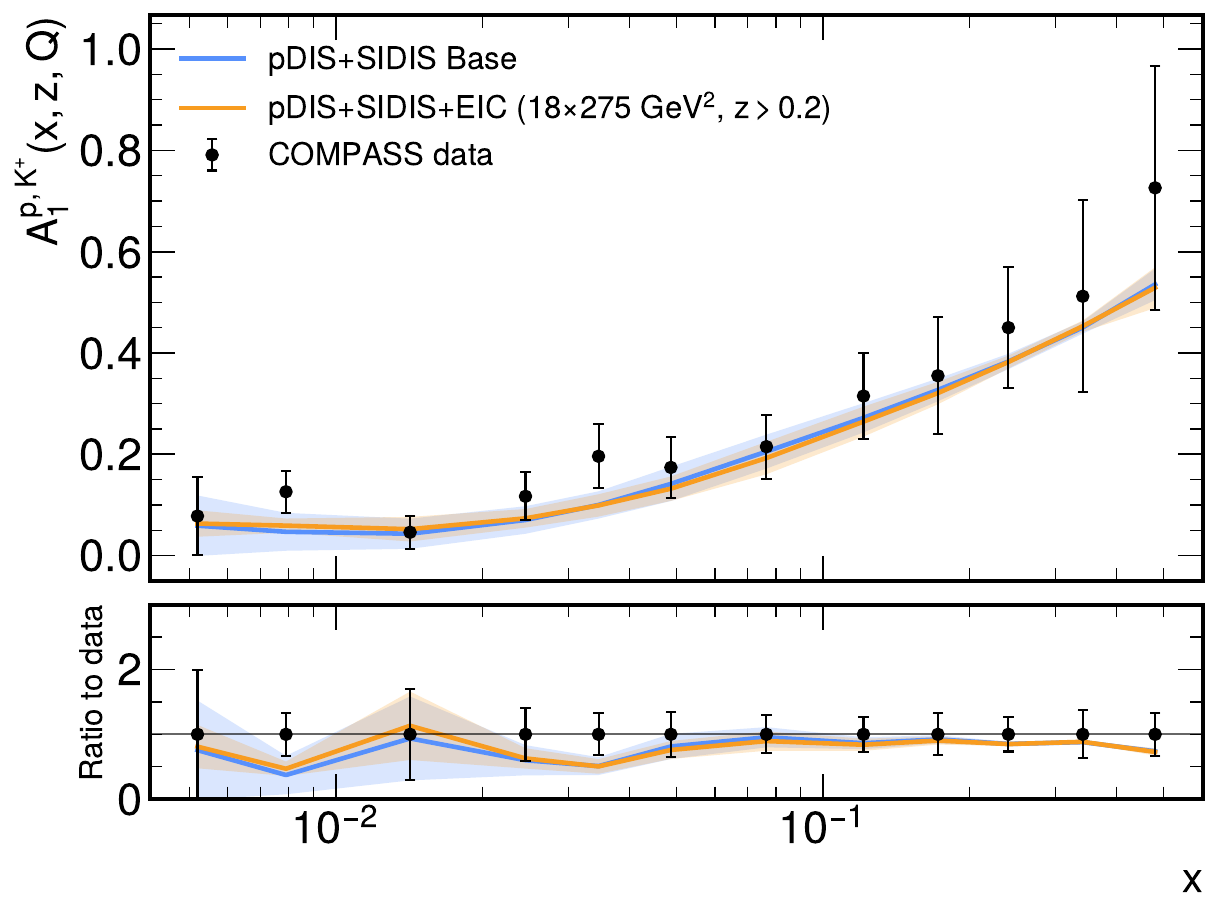}  \\ 
\includegraphics[width=0.48\textwidth]{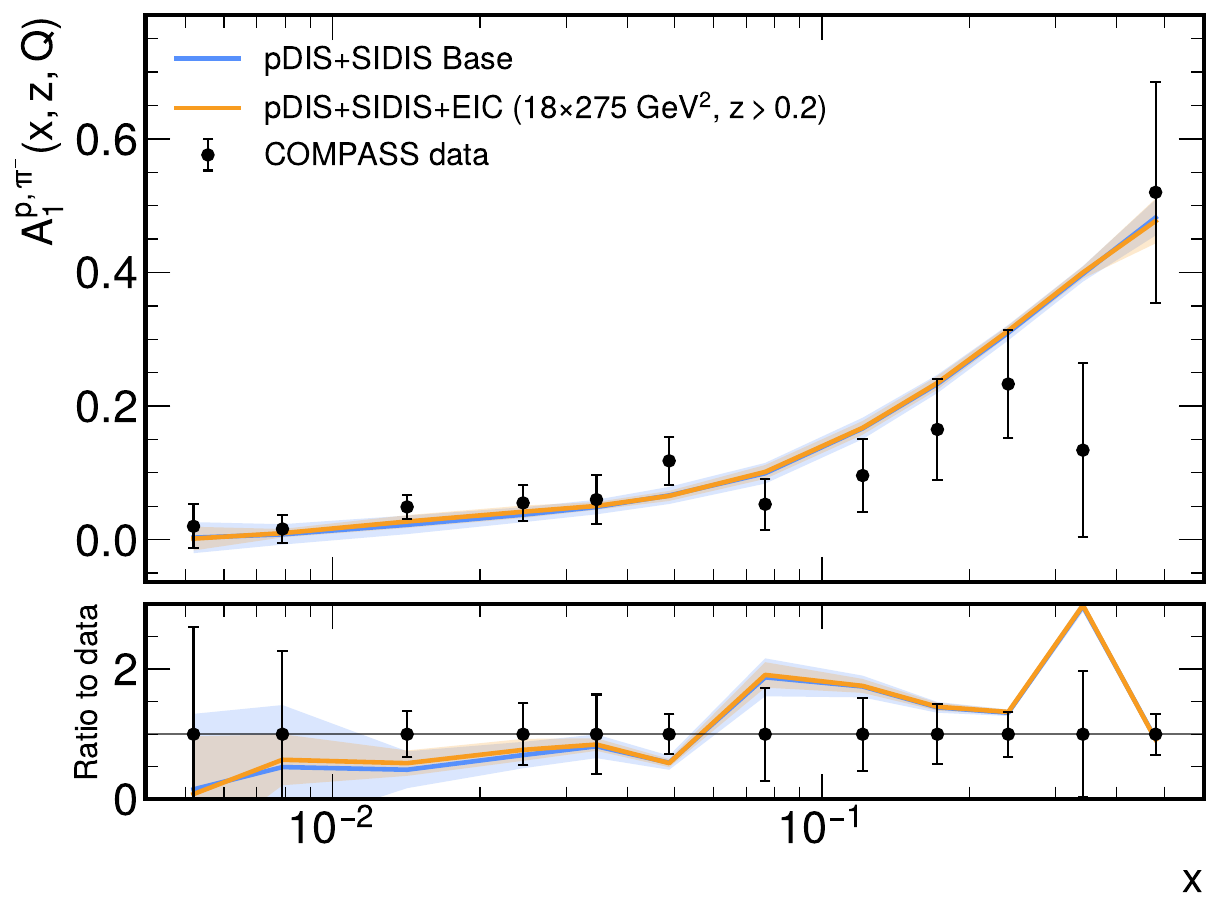} 
\includegraphics[width=0.48\textwidth]{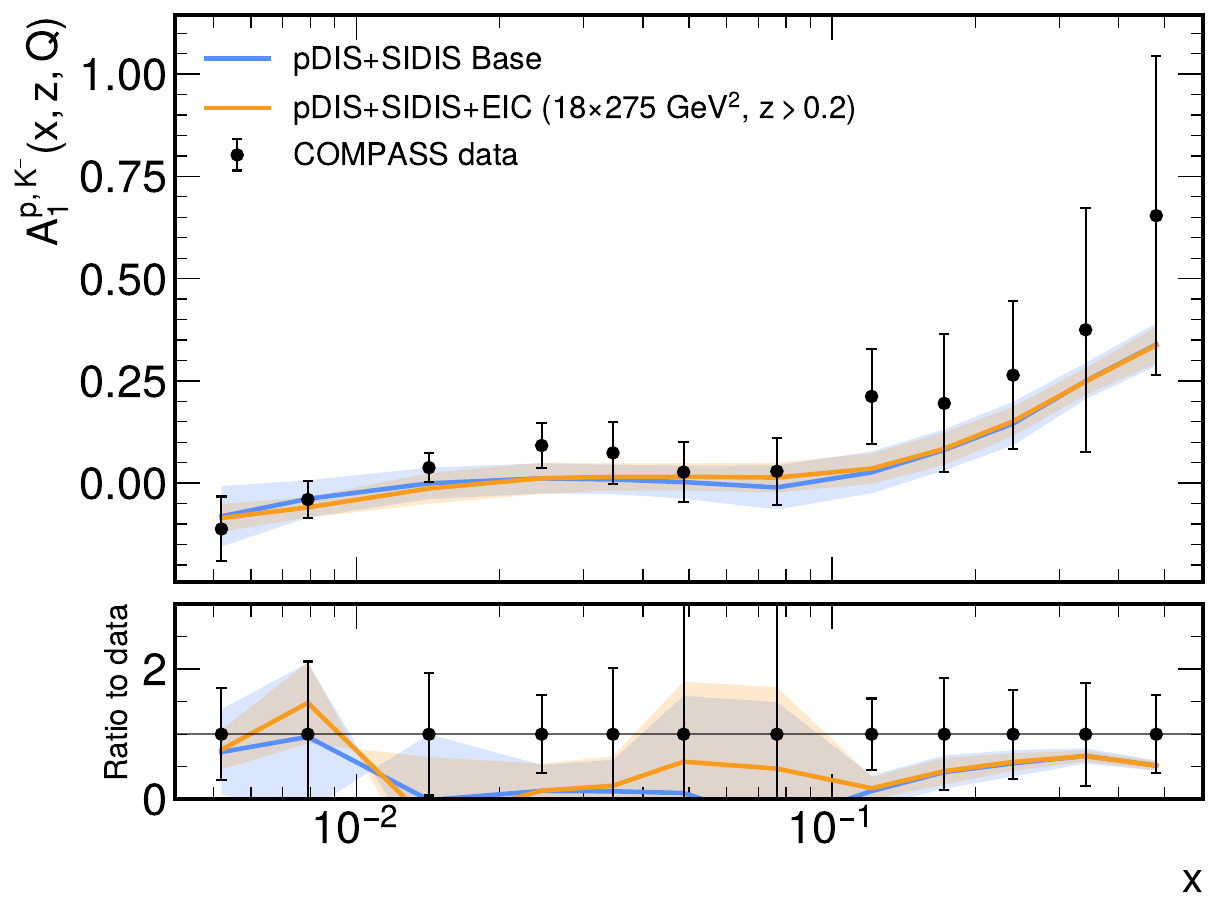} 
\begin{center}
\caption{ \small 
COMPASS measurements of the longitudinal SIDIS double-spin asymmetry $A_1^{p,h^\pm}(x,z,Q)$ for identified hadrons, 
compared to NLO predictions from the {\tt pDIS+SIDIS Base} fit and from the {\tt pDIS+SIDIS+EIC} fit (18$\times$275~GeV$^2$, $z>0.2$).
Shaded bands represent the propagated fit uncertainty.
Lower panels show the ratio theory/data, with the data displayed at unity and their relative experimental uncertainties. }   
\label{fig:compass_hplus_base_vs_eic}
\end{center}
\end{figure*}

To illustrate the impact of the EIC pseudodata on polarized SIDIS phenomenology, we compare
COMPASS measurements of the longitudinal double-spin asymmetry $A_1^{p,h^\pm}$ for identified hadrons
with the corresponding NLO predictions obtained from our global analysis.
Figure~\ref{fig:compass_hplus_base_vs_eic} shows the COMPASS $A_1^{p,\pi^\pm}$ and $A_1^{p,K^\pm}$
data as functions of Bjorken $x$, together with the NLO predictions from the {\tt pDIS+SIDIS Base} fit
and from the {\tt pDIS+SIDIS+EIC} fit (18$\times$275~GeV, $z>0.2$).
The shaded bands represent the propagated fit uncertainty on the theoretical predictions,
evaluated point-by-point at the same kinematics as the data.

Overall, both fits provide a comparable description of the COMPASS asymmetries within uncertainties,
and the inclusion of EIC information does not induce visible distortions of the central predictions
in the fixed-target kinematic region.
The main effect is a clear reduction of the theory uncertainty towards the lowest-$x$ bins,
where present-day constraints are weakest and the baseline fit is more sensitive to small-$x$
extrapolation.
This trend is visible across the pion and kaon channels and is particularly relevant for kaon
asymmetries, where the flavor sensitivity to the strange sea is intrinsically intertwined with the
fragmentation input.
The lower ratio panels further confirm that the changes induced by EIC pseudodata are primarily a
tightening of uncertainties rather than a systematic shift in the central theory/data agreement.

%
%
%
%

\section{Summary and conclusion}\label{summary} 

In this work we have presented a new NLO determination of the helicity-dependent parton distribution
functions of the proton from a global QCD analysis of longitudinally polarized inclusive DIS and
charge-separated SIDIS measurements. The baseline fit is built upon the world fixed-target data set,
including proton, deuteron and neutron information, and uses a consistent leading-twist collinear
factorization framework with NLO DGLAP evolution in the $\overline{\mathrm{MS}}$ scheme.
For SIDIS, the predicted double-spin asymmetries are computed at NLO accuracy through the convolution
of polarized PDFs with NLO coefficient functions and fragmentation functions, and the analysis is
performed within a Monte Carlo replica methodology, providing an ensemble of PDF replicas suitable
for uncertainty propagation.

A central physics goal of this study is to quantify the potential impact of future EIC measurements
on the flavor decomposition of the polarized PDFs and on the poorly constrained small-$x$
behavior of the sea-quark and gluon helicity distribution. To this end we have supplemented the baseline data
set with detector-level EIC SIDIS pseudodata for identified $\pi^+$, $\pi^-$, $K^+$  and $K^-$ production in two
representative beam-energy configurations, $E_e\times E_p = 5\times41~\mathrm{GeV^2}$ and
$18\times275~\mathrm{GeV^2}$, corresponding to complementary kinematic reaches in $(x,Q^2)$. 
The projected EIC measurements provide a substantially extended lever arm in both $x$ and $Q^2$,
including dense coverage in $Q^2$ at fixed $x$, which enhances the scaling-violation information
entering the NLO DGLAP evolution. When combined with charge-separated SIDIS flavor tagging, this
translates into improved constraints on sea-quark helicity combinations (through pion channels for
$\Delta\bar u$ and $\Delta\bar d$, and kaon channels for $\Delta s$) and into a significant reduction
of the uncertainty on $\Delta g$ in the small-$x$ region where current data are largely
extrapolative.

To connect these improvements to the proton spin decomposition, we have examined truncated first
moments of the quark-singlet and gluon helicity distributions as functions of the lower integration
limit $x_{\min}$. The baseline fit exhibits rapidly growing uncertainties as $x_{\min}$ is lowered,
reflecting the limited present constraints at small $x$, especially for $\Delta g$.
With the addition of EIC pseudodata, the uncertainty on the truncated gluon moment is markedly
reduced at small $x_{\min}$, and this improvement propagates directly to the uncertainty of the
combination $\tfrac12\Delta\Sigma+\Delta g$ relevant for the quark-plus-gluon spin contribution.
We find that the EIC-augmented fit yields a visibly improved agreement and reduced theory 
uncertainty towards the lowest-$x$ region covered by the existing 
SIDIS data sets.

Several refinements are natural targets for future work. On the theory side, a more complete
assessment of nonperturbative effects (target-mass and higher-twist corrections at low $Q^2$ and
$W^2$, and nuclear corrections for deuteron data) would sharpen the interpretation of the fixed-target
constraints, particularly given our comparatively loose $W^2$ selection. In addition, the extraction
of sea-quark helicity PDFs from SIDIS remains intrinsically linked to the FFs 
input; a systematic propagation of FFs uncertainties and/or a combined global fit
of polarized PDFs and FFs would further strengthen the robustness of the flavor
separation. 
Finally, extending the perturbative accuracy and/or the phenomenological scope - for
example through higher-order SIDIS corrections, a broader set of polarized processes (such as
 polarized proton-proton data), and/or future treatments beyond collinear factorization - will be
important for fully exploiting the precision program enabled by the EIC.

Overall, our results provide a quantitative demonstration of the transformative potential of EIC
SIDIS measurements for polarized PDF phenomenology. 
By extending the kinematic reach to much smaller
$x$ and providing a multi-scale lever arm in $Q^2$, the EIC enables decisive improvements in the
determination of the gluon helicity at low $x$ and in the flavor decomposition of the polarized sea.
These advances represent a key step towards a more complete and precise understanding of the
spin structure of the proton within QCD.

\section*{Availability of polarized proton PDF sets}\label{LHAPDF}

The polarized PDFs extracted in this analysis are made publicly available
in the standard \texttt{LHAPDF} format~\cite{Buckley:2014ana}.
The released grids contain the full Monte Carlo ensemble of replicas, which encodes the PDF
uncertainties and enables their propagation to any observable of interest.

The sets are provided at NLO accuracy and follow the naming convention
\texttt{pDIS\_SIDIS\_<SETNAME>}, where the label indicates the perturbative accuracy and the inclusion
of the EIC SIDIS pseudodata in the fit.\footnote{We provide separate sets corresponding to the
different fit scenarios discussed in Sec.~\ref{Results}, \emph{e.g.}\ the {\tt pDIS+SIDIS base} fit and
the EIC-augmented fits for the $5\times41$ and $18\times275$ configurations, and for both choice of 
 $z>0.1$ and $z>0.2$.}
Each replica corresponds to an independent fit to a Monte Carlo replica of the input dataset.
Central values are obtained as the average over the replica ensemble, while PDF uncertainties can be
computed using either the standard deviation (for approximately Gaussian distributions) or percentiles. 

The public grid of HAPS-pPDF1.0, documentation, and
plotting/usage examples are hosted in the accompanying GitHub repository~\cite{HAPS-pPDF1.0}. 
Future updates, including additional EIC configurations and refinements of the SIDIS
theory inputs (such as an extended treatment of fragmentation-function uncertainties), will 
be released through the same platform.

\section*{Acknowledgement}   

The authors gratefully acknowledge helpful discussions with Charlotte Van Hulse. 
This work was supported by the School of Particles and Accelerators at the 
Institute for Research in Fundamental Sciences (IPM).  
Hamzeh Khanpour appreciates the financial support from NAWA under grant 
number BPN/ULM/2023/1/00160, as well as from the IDUB program at AGH University.   
The work of  UGM was also supported in part by the Chinese
Academy of Sciences (CAS) President's International
Fellowship Initiative (PIFI) (Grant No.\ 2025PD0022).

\clearpage
\newpage



\end{document}